%% file: template.tex
\documentclass{article}

\usepackage{arxiv}

\usepackage[utf8]{inputenc} % allow utf-8 input
\usepackage[T1]{fontenc}    % use 8-bit T1 fonts
\usepackage{hyperref}       % hyperlinks
\usepackage{url}            % simple URL typesetting
\usepackage{booktabs}       % professional-quality tables
\usepackage{amsfonts}       % blackboard math symbols
\usepackage{nicefrac}       % compact symbols for 1/2, etc.
\usepackage{microtype}      % microtypography
\usepackage{amsmath,amssymb,amsfonts}
\usepackage{cleveref}       % smart cross-referencing
\usepackage{lipsum}         % Can be removed after putting your text content
\usepackage{graphicx}
\usepackage{natbib}
\usepackage{doi}

\usepackage{cite}
\usepackage{algorithmic}
\usepackage{graphicx,color}
\usepackage{textcomp}
\PassOptionsToPackage{hyphens}{url}\usepackage{hyperref}\usepackage[caption=false,font=footnotesize]{subfig}
\usepackage{pifont}
\usepackage{rotating}
\usepackage{array}

\newcommand{\cmark}{\ding{51}} % ✓
\newcommand{\xmark}{\ding{55}} % ✗
\newcommand{\na}{N/A}

\title{HaLert: A Resilient Smart City Architecture for Post-Disaster Based on Wi-Fi HaLow Mesh and SDN}

\newif\ifuniqueAffiliation
\ifuniqueAffiliation % Standard variant of author block
\author{ \href{https://orcid.org/0000-0000-0000-0000}{\includegraphics[scale=0.06]{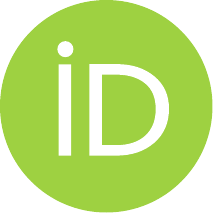}\hspace{1mm}David S.~Hippocampus}\thanks{Use footnote for providing further
		information about author (webpage, alternative
		address)---\emph{not} for acknowledging funding agencies.} \\
	Department of Computer Science\\
	Cranberry-Lemon University\\
	Pittsburgh, PA 15213 \\
	\texttt{hippo@cs.cranberry-lemon.edu} \\
	\And
	\href{https://orcid.org/0000-0000-0000-0000}{\includegraphics[scale=0.06]{orcid.pdf}\hspace{1mm}Elias D.~Striatum} \\
	Department of Electrical Engineering\\
	Mount-Sheikh University\\
	Santa Narimana, Levand \\
	\texttt{stariate@ee.mount-sheikh.edu} \\
}
\else
\usepackage{authblk}

\newbox{\orcid}\sbox{\orcid}{\includegraphics[scale=0.06]{orcid.pdf}} 
\author[1]{%
	\href{https://orcid.org/0009-0001-7529-5857}{\usebox{\orcid}\hspace{1mm}Ana Rita Ortigoso\thanks{\texttt{ana.l.ortigoso@ipleiria.pt}}}%
}
\author[1]{%
	\href{https://orcid.org/0009-0000-2300-8441}{\usebox{\orcid}\hspace{1mm}Gabriel Vieira\thanks{\texttt{gabriel.m.vieira@ipleiria.pt}}}%
}
\author[1]{%
	\href{https://orcid.org/0000-0001-9726-1087}{\usebox{\orcid}\hspace{1mm}Daniel Fuentes\thanks{\texttt{daniel.fuentes@ipleiria.pt}}}%
}
\author[1]{%
	\href{https://orcid.org/0000-0003-2571-7940}{\usebox{\orcid}\hspace{1mm}Luís Frazão \thanks{\texttt{luis.frazao@ipleiria.pt}}}%
}
\author[1]{%
	\href{https://orcid.org/0000-0002-2353-369X}{\usebox{\orcid}\hspace{1mm}Nuno Costa \thanks{\texttt{nuno.costa@ipleiria.pt}}}%
}
\author[1]{%
	\href{https://orcid.org/0000-0001-5062-1241}{\usebox{\orcid}\hspace{1mm}António Pereira \thanks{\texttt{apereira@ipleiria.pt}}}%
}

\affil[1]{Computer Science and Communication Research Centre, Polytechnic University of Leiria, 2411-901 Leiria}

\fi

\renewcommand{\headeright}{Preprint}
\renewcommand{\undertitle}{Preprint}
\renewcommand{\shorttitle}{HaLert}

\hypersetup{
pdftitle={HaLert: A Resilient Smart City Architecture for Post-Disaster Based on Wi-Fi HaLow Mesh and SDN},
pdfsubject={cs.NI, cs.SI, cs.SE, cs.OH},
pdfauthor={Ana Rita Ortigoso, Gabriel Vieira, Daniel Fuentes, Luís Frazão, Nuno Costa e António Pereira},
pdfkeywords={IEEE 802.11ah, IEEE 802.11s, LoRa, LTFH, Mesh, Post-disaster communication, SDN, Smart City, Wi-Fi HaLow},
}

\begin{document}
\maketitle

\begin{abstract}
Events such as catastrophes and disasters are, in most cases, unpredictable. Consequently, reusing existing infrastructures to develop alternative communication strategies after disasters is essential to minimise the impact of these events on the population's ability to communicate and promptly receive alerts from authorities. In this context, the emergence of smart cities, characterised by dense and geographically distributed IoT networks, presents significant potential for such reuse. This work proposes HaLert, a resilient architecture for smart cities based on a Wi-Fi HaLow IEEE 802.11s mesh network, whose resources can be readily reallocated to support a emergency communication system to exchange messages (including text, location, image, audio, and video) between citizens, authorities, and between both parties. To facilitate remote monitoring and configuration of the network, the architecture incorporates the SDN (Software-Defined Networking) paradigm, supported by a LoRa controlled flooding mesh network. A prototype was developed based on this architecture and tested in a real urban scenario comprising both indoor and outdoor environments. The results demonstrated that, despite the significant impact of obstacles, lack of line-of-sight, and terrain slopes on the latency (average latency between 15 and 54.8 ms) and throughput (upload bitrates between 134 and 726 Kbps and
download bitrates between 117 and 682 Kbps) of the Wi-Fi HaLow network, it remained stable and resilient, successfully providing all functionalities associated with the HaLert architecture. The tests conducted on the LoRa network revealed a high average message success rate of 94.96\%.
\end{abstract}

% keywords

\input{Sections/1Introduction}

% RELATED WORK
\input{Sections/2RelatedWork}

% NETWORK ARCHITECTURE
\input{Sections/3NetworkArchitechture}

% DEVELOPED PROTOTYPE
\input{Sections/4DevelopedPrototype}

% TESTS
\input{Sections/5Tests}

% DISCUSSION AND FUTURE WORK
\input{Sections/6Discussion}

% CONCLUSION
\input{Sections/7ConclusionandFutureWork}

\bibliographystyle{unsrtnat}
\bibliography{References/refs.bib}  %%% Uncomment this line and comment out the ``thebibliography'' section below to use the external .bib file (using bibtex) .

\end{document}

%% file: Sections/1Introduction.tex
\section{INTRODUCTION}

Catastrophes and disasters are, in most cases, inherently unpredictable. Consequently, it is important to establish effective systems that enable citizens to maintain communication and receive timely alerts from the authorities. This preparedness is essential to protect public safety and ensure an efficient and coordinated response during emergency situations. Currently, such alerts and official communications are typically disseminated through mobile networks, employing mechanisms such as cell broadcast \citep{Parsons2023} or location-based Short Message Service (SMS) \citep{Aldalahmeh2018} by public safety authorities. Additionally, citizens should be able to easily report incidents to authorities. To reduce anxiety and concern among citizens during such crises, particularly regarding the status and location of their loved ones, maintaining communication between citizens is highly desirable. Mobile networks naturally emerge as a preferred solution due to their widespread coverage and accessibility for contacting authorities and family. However, using mobile networks in a post-disaster context has notable limitations. Historical examples illustrate this vulnerability clearly: approximately 4\,900 base stations became inoperative during Japan’s 2011 earthquake and tsunami \citep{NDTJEO2012}; Haiti’s 2010 earthquake led to the collapse of the capital's mobile infrastructure \citep{ITU2010}; and Hurricane Maria in Puerto Rico (2017) left over 95\% of telecommunication towers out of service \citep{Reuters2017}. These incidents highlight that catastrophes can result in partial or complete failure of mobile networks due to physical damage to infrastructure, prolonged power outages, or transmission system failures. Even partial network failures often lead to service denial due to overload on the remaining operational antennas. Another critical limitation is that mobile antennas cannot function without electricity, and their backup systems typically last less than eight hours \citep{Reuter2014}.

Simultaneously, integrating mobile computing systems within urban data management networks has given rise to the concept of a Smart City \citep{Kirimtat2020}. Despite the absence of a universally accepted definition of a Smart City \citep{Zanella2014, Ismagilova2019}, it is commonly defined as a city employing technological solutions to enhance urban management and efficiency, benefiting both residents and businesses \citep{ComissaoEuropeia_SmartCity}.

The primary aim of smart cities is optimising public resource utilisation, enhancing service quality for citizens, and reducing operational costs for local administrations \citep{Zanella2014}. This includes reducing emissions, implementing intelligent urban transport systems, enhancing water and waste management through smart technologies, improving energy efficiency in buildings, fostering interactive city governance, ensuring public safety, and addressing the specific needs of the eldery population \citep{ComissaoEuropeia_SmartCity}.

Framed by this perspective, there is an opportunity to enhance disaster management the efficiency in smart cities by repurposing existing Internet of Things (IoT) infrastructures to mitigate disaster impacts. A practical example is observed in Florida, frequently affected by hurricanes, where mesh Wi-SUN networks used by smart meters can identify non-functional network segments and notify energy providers. In some scenarios, the network directly communicates with energy distribution controllers, reconfiguring electrical networks to mitigate infrastructure damage \citep{Beecher_IotForAll_023}.

Nevertheless, Wi-SUN technology has data rate limitations (50 Kbps to 2.4 Mbps) \citep{Tian2021}, which restrict high-volume IoT data transmission, such as real-time video. In contrast, Wi-Fi HaLow (IEEE 802.11ah), officially standardised in 2016 \citep{Tian2021}, presents significant advantages over Wi-SUN, offering higher data rates (150 kbps to 86.7 Mbps) \citep{WFA2020} and extended battery life \citep{WFA2021}. This increased battery life is particularly beneficial in emergency scenarios where battery-powered devices are advantageous. Wi-Fi HaLow is also suitable for emergencies and IoT contexts due to its long-range capability (up to 1 km), Internet Protocol (IP) connectivity \citep{WFA2020}, strong penetration through obstacles facilitated by operation at sub-1 GHz frequencies (863–868 MHz in Europe, 902–928 MHz in North America, 755–787 MHz in China \citep{Tian2021}, and 915–928 MHz in Australia and New Zealand \citep{WFA2021}), and narrow bandwidth channels derived from IEEE 802.11ac standards with a tenfold reduced clock frequency \citep{Tian2021, Lee2021}. IEEE 802.11ah supports channel bandwidths of 1, 2, 4, 8, and 16 MHz, with 1 MHz and 2 MHz channels mandatory for compliance, and can connect up to 8191 stations per access point \citep{Muteba2019}.

Considering these factors, this paper presents HaLert, a resilient wireless IoT architecture designed for smart cities, whose resources can be dynamically reallocated to be used as a emergency communciation system to ensure effective communication between citizens and authorities in post-disaster scenarios. HaLert utilises Software-Defined Networking (SDN), which logically separates the network control and data planes, enabling centralised, programmable, and flexible network management \citep{ONF_SDN_definition}. This flexibility significantly contributes to network resilience and responsiveness during emergencies.

To prevent IoT network failures from affecting device management, monitoring, and configuration capabilities, SDN should be deployed on an autonomous network based on a Long Range (LoRa) mesh network. This tecnology and topology was explicitly selected for its inherent resilience and suitability in emergency contexts. Despite LoRa's limited data rates (300 bps to 50 Kbps) \citep{Chilamkurthy2022}, its proprietary Chirp Spread Spectrum (CSS) modulation technique \citep{Devalal2018, AyoubKamal2023, Sallum2020, GAITAN2020} ensures robust, long-range communication with minimal power requirements. Originally developed in the 1940s \citep{AyoubKamal2023} for military and space applications, CSS modulation significantly improves network reliability and resilience in challenging post-disaster environments.

Integrating SDN with LoRa and mesh topology enables remote configuration of numerous devices and incorporates redundancy mechanisms, enhacing overall network resilience.

\input{Sections/1Introduction/Contributions}

\input{Sections/1Introduction/Objectives}

\input{Sections/1Introduction/PaperOrganisation}

%% file: Sections/1Introduction/Contributions.tex
\subsection{CONTRIBUTIONS}

This work presents several contributions towards the development of resilient and adaptable network infrastructures in urban environments, with a particular focus on emergency situations and disaster response. The main contributions are as follows:

\begin{itemize}
    \item The design of a resilient and flexible architecture for Smart Cities, enabling the dynamic reallocation of computational resources via SDN to support emergency communications.
    \item Proposal of a novel application of Wi-Fi HaLow for post-disaster response scenarios.
    \item Real-world evaluation of an IEEE 802.11s Wi-Fi HaLow mesh network in a mixed urban indoor/outdoor environment.
    \item A new aproach to the LoRa Temporal Frequency Hopping (LTFH) mechanism \citep{Ortigoso2024} for network configuration and monitoring via SDN, including optimization of Google's Protocol Buffers (Protobuf) messages, achieving a reduction of 50\% in payload size.
\end{itemize}

%% file: Sections/1Introduction/Objectives.tex
\subsection{OBJECTIVES}

The primary objective of this work is to design and evaluate a resilient smart city architecture to ensure reliable post-disaster communication between citizens and authorities, as well as among citizens themselves. The following specific objectives were defined to accomplish this goal:

\begin{itemize}
\item Design a dual-mesh network architecture integrating IEEE 802.11s Wi-Fi HaLow and LoRa technologies, tailored specifically for urban emergency scenarios.
\item Develop an SDN platform enabling remote configuration and management of IoT infrastructure via a LoRa-based controlled flooding mesh network.
\item Implement emergency communication tools, including captive portals and dedicated chat applications, providing differentiated services for citizens and emergency authorities.
\item Optimise LoRa-based SDN message formats utilising Protocol Buffers, reducing payload sizes to enhance message transmission speed, increase information throughput, and minimise radio spectrum occupancy, thus decreasing interference with concurrent LoRa and other communications operating on similar frequencies.
\item Validate the proposed architecture through prototype development and practical evaluation in a realistic urban environment comprising both indoor and outdoor scenarios.
\item Assess the performance and limitations of the Wi-Fi HaLow mesh network under constrained conditions, specifically analysing latency, throughput, and signal stability.
\item Evaluate the reliability and efficiency of the LoRa mesh network for SDN control messages, focusing on message success rates and transmission redundancy.
\end{itemize}

%% file: Sections/1Introduction/PaperOrganisation.tex
\subsection{PAPER ORGANISATION}

The remainder of this paper is organised as follows. Section II provides a review of existing studies focused on emergency communication systems utilising LoRa and Wi-Fi HaLow technologies, including a comparative summary of these studies presented in \autoref{tab:comparison_emergency_communication}. Section III introduces the HaLert architecture, detailing its dual-mesh network structure and integration within smart city environments. Section IV describes the developed prototype, detailing the implementation of the SDN platform, IoT platform and emergency communication chats. Section V presents the experimental methodology and associated results, covering functional and performance evaluations conducted in realistic urban settings. Section VI critically discusses the identified limitations, challenges encountered, and potential avenues for scaling and further optimisation. Finally, Section VII concludes by summarising the primary contributions and outlining directions for future research.

%% file: Sections/2RelatedWork.tex
\section{RELATED WORK}

Multiple studies have explored off-grid and decentralised communication systems based on LoRa technology for emergency scenarios. In \citep{Macaraeg2020} a decentralised LoRa-based off-grid mesh communication system for emergencies was developed, enabling smartphones without native LoRa capabilities to connect via Bluetooth. The LoRa mesh network employs the Ad hoc On-demand Distance Vector (AODV) routing algorithm. Similarly, \citep{Dalpathadu2021} proposed a solution based on the commercial LoRaMAC system to facilitate post-disaster operations by communicating sensor data to rescuers and external parties using opportunistic networks and epidemic forwarding.

\citep{Ranasinghe2024} proposed a LoRa multi-hop network as an alternative to the public Internet for Early Warning of Earthquakes (EEW). Practical demonstrations using Lilygo LoRa32 devices achieved the dissemination of warnings across a 30 km urban area in 2.48 seconds.

In this work \citep{Vithayathil2021}, the authors presented an offline communication system for disaster rescue operations that allows citizens to communicate with rescue teams. The system employs LoRa for communication and Wi-Fi for victim interface, allowing them to submit rescue-relevant information via a captive portal. These messages are transmitted via MQTT to the Adafruit IoT cloud platform, while emergency notifications are also sent via email to rescue teams. The system was tested using Heltec Wi-Fi LoRa 32 devices. \citep{Stiballe2023} demonstrated a system with a similar approach but employing a routing mechanism inspired by Dynamic Source Routing, appending the forwarding node Identifier (ID) to each relayed message. Demonstrations utilised Raspberry Pi 3B devices with SX1262 868M LoRa Hardware Attached on Top (HAT). Likewise, \citep{PueyoCentelles2021} presented an emergency communication system specifically designed to enable civilians to safely exchange information rather than between citizens and authorities regarding their security status in critical situations. Based on LoRaWAN technology, the system requires pre-installed home devices that interact with users via Wi-Fi or Bluetooth. The user interface is provided through an embedded web server on the device or via a mobile application, allowing users to send short messages. Additionally, the system facilitates secure and effective civilian communication while providing valuable data to emergency teams, thus enhancing the efficiency of crisis response.

Other studies focused on the evaluation of Wi-Fi HaLow in post-disaster scenarios. \citep{Khan2017,Khan2018a, Khan2018b, Khan2018} investigated signal propagation models, radar-based Wi-Fi methods, and simulation techniques for collapsed infrastructure environments. \citep{Khan2019} conducted field studies in Pakistan, showing Wi-Fi HaLow's superior performance compared to conventional 2.4 GHz and 5 GHz frequencies due to reduced attenuation in debris-affected areas, using a modified path loss model (PL-Collapsed). This conclusion was also achieved in \citep{Khan2017} through simulations incorporating a generalised debris model. \citep{Khan2020} highlighted the superior penetration capacity of the 779 MHz frequency in studies conducted in China, optimising coverage attributes using Shannon entropy and the TOPSIS method. Additionally, \citep{Khan2019a} explored its application in post-disaster IoT for survivor localisation and breath detection.

\citep{Khorov2019} proposed a mathematical model for heterogeneous Wi-Fi HaLow networks, differentiating between user stations generating saturated traffic and emergency sensors producing non-saturated traffic. Using the Enhanced Distributed Channel Access mechanism, the study determined the optimal number of devices and sensors for efficient and reliable emergency alert transmission. \citep{Purat2022} introduced HaLowNet, a Wi-Fi HaLow-based information system designed for emergency medical scenarios lacking Information Technology (IT) infrastructure. HaLowNet connects network nodes across locations via Wi-Fi HaLow and IEEE 802.11 b/g/n for intra-location device connectivity. It provides a repository for emergency applications accessible without internet connectivity.

\citep{Thangadorai2025} proposed and evaluated an IEEE 802.11s Wi-Fi HaLow-based mesh network using the Better Approach to Mobile Ad-hoc Networking advanced (batman-adv) routing protocol for human-to-human communication in indoor and outdoor scenarios. The system achieved video calls over 800 m and voice calls over 1 km with throughputs between 700 kbps and 420 kbps, respectively. \citep{Chounos2023} assessed the practical use of Unmanned Aerial Vehicles (UAV) with IEEE 802.11ah for emergency scenarios. The study found that at distances of 0 to 1000 m, the average throughput for a 1 MHz channel was 1.69 Mbps at altitudes of 35 m and 70 m, while a 2 MHz channel yielded 2.61 Mbps and 2.56 Mbps at the same altitudes. \citep{Riza2020} proposed an IEEE 802.11ah-based architecture for connecting medical sensors, robots, and hospital devices, enabling remote patient monitoring for Covid-19 cases. Data transmission was facilitated via 3G/4G networks to ensure continuous medical supervision.

The article \citep{Kane2023} conducted a real-world outdoor evaluation of Wi-Fi HaLow's performance relative to traditional Wi-Fi and LoRa, discussing its applicability in smart grid scenarios. The assessment involved various distances (between 50 and 1000 m) between an access point and an end device (station) in a scenario combining line-of-sight with moderate environmental obstructions, such as trees, analysing throughput, latency, and reliability metrics.

\citep{Urbanelli2024} introduced ERMES Chatbot, a mobile-based conversational system facilitating real-time bidirectional communication between citizens and emergency authorities via Telegram. The system's emergency response interface allows authorities to share real-time locations, report field conditions with detailed measurements and damage assessments, and receive instructions from control rooms. The citizen interface enables users to report emergencies, share real-time information, receive broadcast communications, and access relevant safety guidelines, supporting text, location, video, image, and audio transmission.

Table \ref{tab:comparison_emergency_communication} provides a comparative analysis of previous works proposing communication systems designed for emergency scenarios. The evaluation considers the communication technology employed and its corresponding network layer, the data types supported (e.g., text, geolocation, multimedia, sensor data), and whether the solutions reuse existing infrastructures to simplify deployment, improve convenience, and reduce costs. Additionally, compatibility with common end-user devices, such as smartphones or laptops, is assessed to ensure widespread accessibility. The implementation of mesh topology is also examined due to its capacity to enhance resilience and network coverage. Furthermore, the availability of user-interaction mechanisms, including captive portals and web or installable applications, is analysed. Finally, the supported communication flows—authority-to-citizen, citizen-to-authority, citizen-to-citizen, and authority-to-authority—are identified and compared.

Most reviewed solutions adopt LoRa-based communication technologies, predominantly transmitting text, geolocation, or sensor data. These systems generally exhibit limited interfacing capabilities and minimal reuse of existing infrastructures. Few studies, notably \citep{Stiballe2023} and \citep{PueyoCentelles2021}, support multiple communication flows and include enhanced user interface features such as captive portals or web applications. It is noteworthy, however, that using LoRaWAN technology can be disadvantageous as it restricts the duty cycle according to gateway specifications, whereas solutions employing LoRa without the Wide-Area Network (WAN) layer allow duty cycle regulations to be bypassed if necessary in critical emergency situations. Conversely, solutions employing Wi-Fi HaLow, such as those presented by \citep{Purat2022} and \citep{Riza2020}, exhibit comparatively limited capabilities in user interaction and supported communication directions, and notably are not designed for direct communication between authorities and/or citizens.

The system proposed by \citep{Urbanelli2024} notably distinguishes itself by leveraging existing infrastructure and supporting diverse media types through a Telegram chatbot, effectively covering multiple communication flows, though it does not propose a specific network architecture. 

In comparison, the proposed solution distinguishes itself by integrating both Wi-Fi HaLow and LoRa technologies in a dual-mesh architecture, providing a resilient and versatile framework specifically designed for post-disaster communication. Unlike the approach proposed in \citep{Urbanelli2024}, which relies on a existing messaging platform, Telegram, without proposing a dedicated network infrastructure, HaLert delivers a comprehensive and autonomous system encompassing physical network design, SDN-based remote management, and application-layer services. A key limitation of Urbanelli’s solution lies in its dependence on infrastructure-based Wi-Fi networks and conventional mobile networks, including domestic, workplace, and public access points, which are highly vulnerable to disruption in emergency contexts. This reliance may result in partial or complete service failure depending on the severity and geographical scope of the disaster. Moreover, by depending entirely on a third-party application (Telegram) for all communication, the solution introduces a single point of failure (SPOF)—any widespread outage or restriction of the platform can render the entire system inoperative. In contrast, HaLert is fully self-contained, enabling resilient communication even in the absence of Internet connectivity. It supports a broad range of data types (text, geolocation, images, audio, video), provides multiple user access methods (captive portal, web app, and PWA), and enables full bidirectional communication between authorities and citizens, as well as among citizens and authorities. This level of integration, autonomy, and fault tolerance positions HaLert as the most robust and comprehensive solution among the systems reviewed.

\begin{table*}[ht]
\centering
\caption{Comparison of communication systems for emergency scenarios}
\scriptsize
\setlength{\tabcolsep}{2pt}
\renewcommand{\arraystretch}{1.2}
\begin{tabular}{
|p{1.0cm} % Article
|p{1.8cm} % Tech (with OSI layer)
|p{2.3cm} % Data type
|p{1.3cm} % Reuses infra
|p{1.2cm} % Interfacing
|p{0.6cm} % Mesh
|p{0.8cm} % Captive portal
|p{0.5cm} % Web App
|p{0.5cm} % Application
|p{1.2cm} % A→C
|p{1.2cm} % C→A
|p{0.9cm} % C↔C
|p{1.2cm}|} % A↔A
\hline
\textbf{Article} &
\textbf{Communication Technology (Layer)} &
\textbf{Data Type} &
\textbf{Reuses Existing Infrastructures} &
\textbf{Interfacing with Existing Devices} &
\textbf{Mesh} &
\textbf{Captive Portal} &
\textbf{Web App} &
\textbf{App} &
\textbf{Authorities to Citizens} &
\textbf{Citizens to Authorities} &
\textbf{Citizens to Citizens} &
\textbf{Authorities to Authorities} \\ \hline

\citep{Macaraeg2020} &
LoRa (PHY/MAC) &
Text and geolocation &
\xmark & \cmark & \cmark & \xmark & \xmark & \xmark & \xmark & \xmark & \cmark & \xmark \\ \hline

\citep{Dalpathadu2021} &
LoRaMAC (MAC) &
Sensor data and geolocation &
\xmark & \xmark & \cmark & \xmark & \xmark & \xmark & \na & \na & \na & \na \\ \hline

\citep{Ranasinghe2024} &
LoRa (PHY/MAC) &
Earthquake Early Warnings &
\xmark & \xmark & \cmark & \xmark & \xmark & \xmark & \na & \na & \na & \na \\ \hline

\citep{Vithayathil2021} &
LoRa (PHY/MAC) &
Text &
\xmark & \cmark & \xmark & \cmark & \cmark & \xmark & \xmark & \cmark & \xmark & \xmark \\ \hline

\citep{Stiballe2023} &
LoRa (PHY/MAC) &
Text and control data &
\xmark & \cmark & \cmark & \cmark & \cmark & \xmark & \cmark & \cmark & \cmark & \xmark \\ \hline

\citep{PueyoCentelles2021} &
LoRaWAN (MAC/Network) &
Text &
\xmark & \cmark & \cmark & \xmark & \cmark & \cmark & \cmark & \cmark & \xmark & \xmark \\ \hline

\citep{Purat2022} &
Wi-Fi HaLow (PHY/MAC) &
Medical, logistics information and network data &
\xmark & \cmark & \xmark & \xmark & \cmark & \xmark & \na & \na & \na & \na \\ \hline

\citep{Chounos2023} &
Wi-Fi HaLow (PHY/MAC) &
Control and telemetry data &
\xmark & \xmark & \xmark & \xmark & \xmark & \xmark & \na & \na & \na & \na \\ \hline

\citep{Riza2020} &
Wi-Fi HaLow (PHY/MAC) &
Medical information and control data &
\xmark & \xmark & \xmark & \xmark & \cmark & \xmark & \na & \na & \na & \na \\ \hline

\citep{Urbanelli2024} &
Telegram chatbot (Application) &
Text, image, video, audio and geolocation &
\cmark & \cmark & \na & \na & \cmark & \na & \cmark & \cmark & \cmark & \cmark \\ \hline

Proposed architecture &
Wi-Fi HaLow (PHY/MAC) + LoRa (PHY/MAC) &
Text, image, video, audio, geolocation and network management data &
\cmark & \cmark & \cmark & \cmark & \cmark & \cmark & \cmark & \cmark & \cmark & \cmark \\ \hline

\end{tabular}
\label{tab:comparison_emergency_communication}
\end{table*}

%% file: Sections/3NetworkArchitechture.tex
\section{NETWORK ARCHITECTURE}

This work proposes an architecture for a wireless emergency network that can be integrated into smart cities. The objective of this network is to leverage existing smart cities infrastructure to ensure communication in post-disaster scenarios, both among citizens, among authorities, and across citizen-authority interactions.

\begin{figure*}
\centerline{\includegraphics[width=\textwidth]{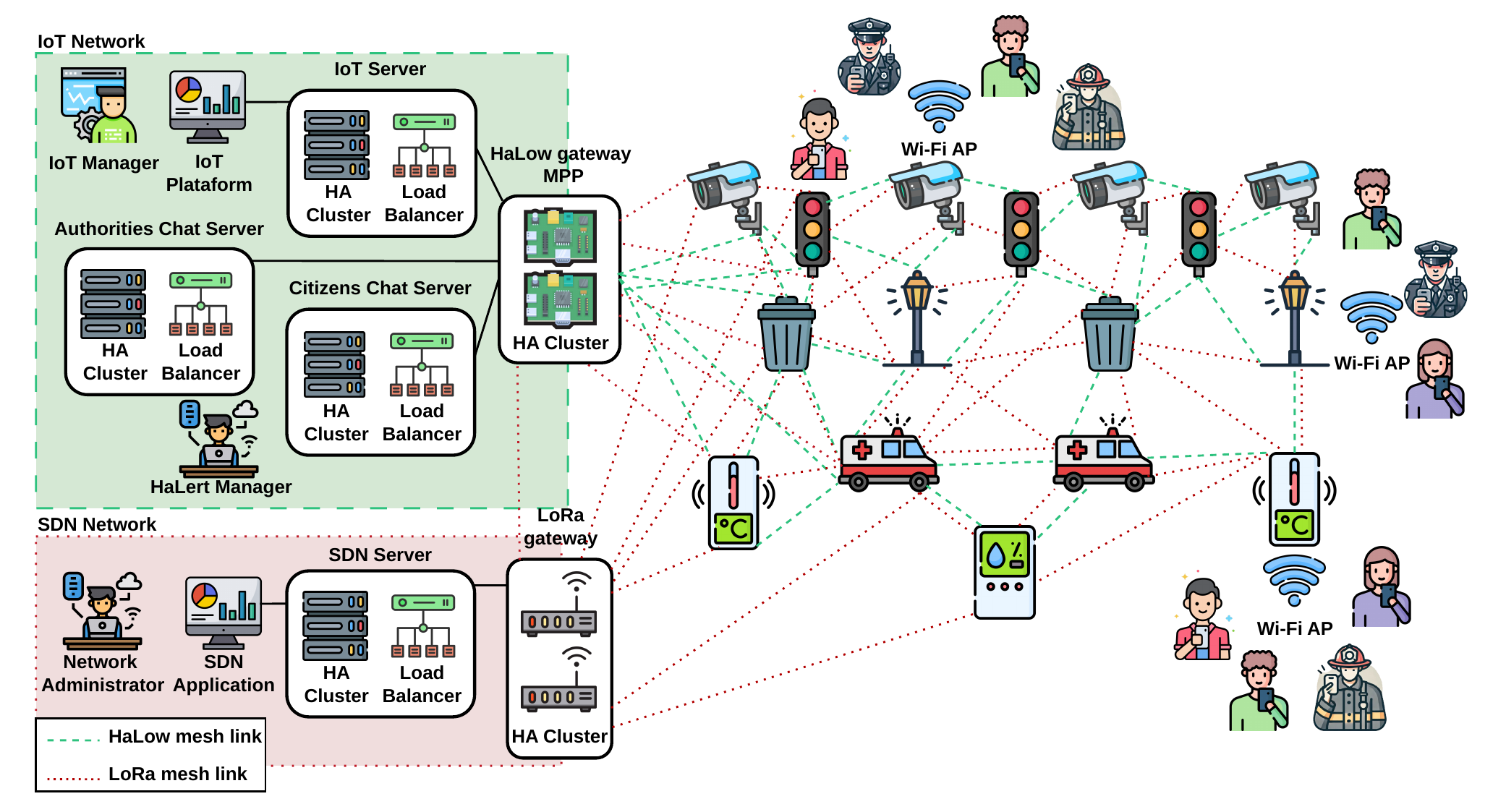}}
\caption{HaLert network architecture. The system comprises two main network segments: the SDN network (shaded in red) and the IoT network (shaded in green). Two types of mesh links interconnect IoT devices: HaLow mesh links (depicted with green dashed lines) and LoRa mesh links (depicted with red dotted lines).\label{fig:net_arch}}
\end{figure*}

As illustrated in \autoref{fig:net_arch}, the proposed architecture is organised into three logical segments: the IoT network, the SDN network, and the urban smart environment. The IoT and SDN networks assume distinct but complementary roles, while the smart environment comprises the operational domain in which these systems interact with real-world devices and end-users.

The IoT network, shaded in green, is responsible for sensing, actuation, and communication services. It includes three core servers: the IoT Server, the Citizens Chat Server, and the Authorities Chat Server. The IoT Server manages the flow of data between the IoT platform and the deployed devices, supporting the integration of environmental and infrastructure sensors and actuators. It also provides a real-time dashboard that enables the IoT manager to monitor the operational status of all devices. The Citizens Chat Server facilitates emergency communication among civilians and the reception of alerts from authorities, while the Authorities Chat Server provides a dedicated channel for institutional communication among emergency responders, law enforcement, and municipal authorities. All IoT servers connect to a HaLow Gateway, deployed in a high availability (HA) cluster, which serves as the Mesh Point Portal (MPP) for the IEEE 802.11s mesh network, enabling the extension of the IoT network over a resilient Wi-Fi HaLow backbone. The HaLert manager is specifically responsible for managing emergency chat servers and validating alerts sent to citizens, ensuring accurate and timely dissemination of critical information.

The SDN network, shaded in red, provides centralised monitoring and control capabilities over the entire infrastructure. It comprises the SDN Server, responsible for maintaining the global network state and applying control rules, and the SDN Application, which provides a management interface operated by the network administrator. This segment adopts the SDN paradigm, which enables logically centralised and remote execution of network operations, including the activation and deactivation of sensors and actuators, enabling or disabling Wi-Fi connectivity, reconfiguring communication links and see the operational status of the devices. The SDN network communicates with the smart city devices via a LoRa mesh network that employs a controlled flooding mechanism to distribute configuration updates. A dedicated LoRa Gateway, deployed in an HA cluster, handles the transmission of control messages and the collection of device status reports. To comply with European duty cycle constraints, typically limited to 1\%, the LTFH mechanism \citep{Ortigoso2024} is employed to optimise airtime efficiency and minimise interference.

To ensure operational resilience and scalability, all servers in both the IoT and SDN networks are deployed with high availability and load balancing mechanisms, preventing single points of failure and ensuring service continuity under adverse conditions.

The urban smart environment includes a wide range of interconnected devices, such as temperature and humidity sensors, surveillance cameras, traffic lights, street lighting, waste bins, and emergency service vehicles. These devices are simultaneously connected to both the IoT and SDN networks. Inter-device communication and data dissemination are achieved via two complementary mesh topologies: a Wi-Fi HaLow mesh (represented by green dashed lines) supports sensing data exchange and access to chat services, while a LoRa mesh (depicted with red dotted lines) provides a control channel for configuration and status reporting. Many of these devices also function as Wi-Fi access points, allowing nearby end-user devices such as smartphones and laptops to connect to emergency chat services. In end-user nodes operated by public authorities, internet access may be selectively enabled to prevent network overload caused by citizen traffic during critical events.

The architecture supports dynamic adaptation to evolving conditions through the SDN paradigm, which allows the network administrator to perform remote reconfigurations in a logically centralised manner. In the event of a disaster, the network can be partially or fully restructured to prioritise connectivity in affected areas. For instance, resources can be reallocated to reinforce degraded regions, while unaffected sectors remain operational. Moreover, additional nodes can be deployed in both the HaLow and LoRa mesh networks to extend coverage and improve fault tolerance.

%% file: Sections/4DevelopedPrototype.tex
\section{DEVELOPED PROTOTYPE}

\begin{figure}
\centerline{\includegraphics[width=\linewidth]{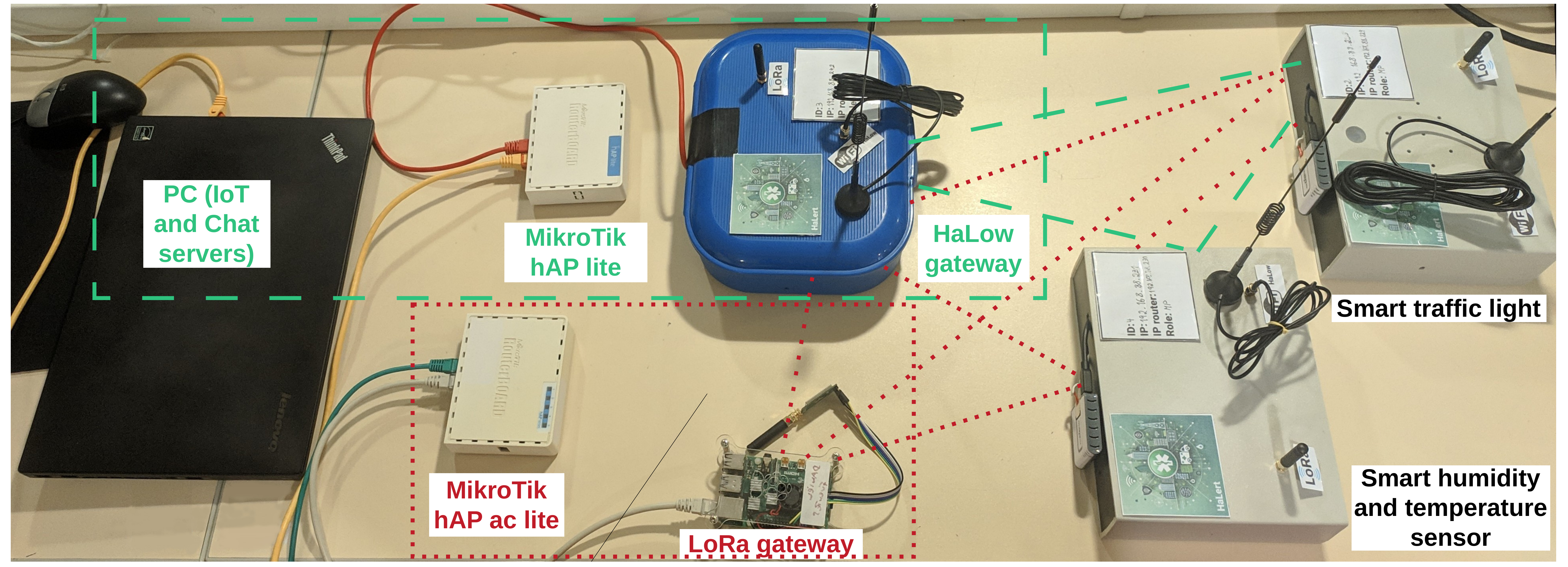}}
\caption{Developed prototype based on HaLert architecutre.
\label{fig:photo_prototype}}
\end{figure}

Based on the proposed network architecture, the prototype presented in \autoref{fig:photo_prototype} was developed to validate and evaluate the functionality of the network and application logic.

Prototypes of IoT devices were implemented for this smart city architecture, specifically a smart temperature and humidity sensor and a smart traffic light. These devices were built using Raspberry Pi 3B boards. For connectivity, HaLow modules AHMB7292S and AHPI7292S were employed—both based on the NRC7292 chipset, differing only in form factor (mikroBUS and HAT, respectively). LoRa connectivity was enabled using E220-900T22D modules. To ensure network access for citizens and authorities in emergency situations, each device was also connected to a MikroTik mAP lite router. The temperature and humidity sensor used was the DHT11, while the traffic light was simulated using an RGB LED. These devices were configured as Mesh Points (MP) in the HaLow network.

Prototypes for both HaLow and LoRa gateways were also developed.

The HaLow gateway used the same hardware as the IoT devices (Raspberry Pi 3B, AHPI7292S, and E220-900T22D), operating as the MPP in the HaLow network and capable of being reconfigured via the LoRa network. Additionally, to facilitate IoT data collection, a Message Queuing Telemetry Transport (MQTT) broker was deployed on this device. 

The LoRa gateway was based on a Raspberry Pi 4 and an E220-900T22D module and functioned as the server for all SDN-related applications.

To prevent frequency collisions, reducing interference and enhancing communication robustness, the HaLow radios were configured to operate using 1 MHz channels centred at 863.5 MHz, while the LoRa radios were configured with the LTFH mechanism, enabling dynamic switching between frequencies of 865, 866, 867, and 868 MHz.

The IoT network routing was handled by a MikroTik hAP lite router, while the SDN network used a MikroTik hAP ac lite router. All other applications, including chat services for citizens and authorities and the IoT application frontend, were hosted on a personal computer.

\input{Sections/4DevelopedPrototype/SDNPlatform}

\input{Sections/4DevelopedPrototype/IoTPlatform}

\input{Sections/4DevelopedPrototype/EmergencyChats}

%% file: Sections/4DevelopedPrototype/SDNPlatform.tex
\subsection{SDN PLATFORM}

The SDN platform aims to enable remote configuration and monitoring of various IoT devices deployed within the smart city environment.

Devices are registered on the platform by providing a predefined set of attributes, as detailed in \autoref{tab:sdn-device-registration}. This information is essential for accurate device identification, geospatial mapping, network role assignment, and contextual management within the SDN platform. Each device is assigned a `Device Name', which serves as a human-readable identifier to simplify interaction through the platform's user interface. The `Device ID' is a unique numeric code used specifically to identify each device within the LoRa network. The `Sensor Type' attribute specifies the type of sensor integrated into the device (e.g., temperature, humidity), enabling the system to apply relevant monitoring and control protocols. The `Coordinates' attribute indicates the device's physical location, which may be manually entered, searched by address, or selected directly via an interactive map, thereby facilitating precise geospatial management and visualisation. The optional Notes' field allows the inclusion of additional pertinent information, such as installation conditions or operational constraints. Lastly, the `Mesh Role' defines the device's specific function within the mesh network—either as a MP or MPP—which determines the configuration options displayed in the user interface according to each device type.

\begin{table}[htbp]
\caption{Information required for device registration.}
\centering
\begin{tabular}{|l|p{10cm}|}
\hline
\textbf{Field} & \textbf{Description} \\
\hline
Device Name & Name assigned to the device \\
\hline
Device ID & Unique identifier of the device \\
\hline
Sensor Type & Type of sensor associated with the device \\
\hline
Coordinates & Location (entered manually, searched by address, or selected on the map) \\
\hline
Notes & Optional field for additional comments \\
\hline
Mesh Role & Role in the mesh network (MPP or MP) \\
\hline
\end{tabular}
\label{tab:sdn-device-registration}
\end{table}

Once registered, devices can be selected either through a table or directly on a map for configuration.

The SDN platform supports a comprehensive set of remote management and diagnostic operations, as detailed in \autoref{tab:sdn-functions}. These include enabling or disabling sensor and Wi-Fi interfaces on individual devices, allowing their functionality to be dynamically adjusted according to the operational context. Devices are also capable of switching between sensing and access point modes, or operating both modes concurrently when required, which increases deployment flexibility in scenarios with high communication and sensing demands. Remote reboot functionality is supported to restore devices in case of malfunction or configuration updates. The platform further enables continuous monitoring by querying the operational status of sensors and access points. It also retrieves the number of client devices currently connected to each access point, providing visibility over network load distribution. Lastly, it allows verification of device connectivity to the SDN controller, supporting timely identification of communication failures. Collectively, these capabilities enable centralised, real-time management and enhance the resilience and adaptability of the system.

\begin{table}[htbp]
\caption{Functions supported by the SDN platform}
\centering
\begin{tabular}{|l|p{7.3cm}|}
\hline
\textbf{Function} & \textbf{Description} \\
\hline
Sensor Control & Activation and deactivation of sensors \\
\hline
Wi-Fi Control & Activation and deactivation of Wi-Fi \\
\hline
Sensor-to-AP Switch & Deactivates sensors and activates AP \\
\hline
AP-to-Sensor Switch & Deactivates AP and activates sensors \\
\hline
Device Reboot & Remotely reboots the device \\
\hline
Sensor Status Check & Verifies whether sensors are active or inactive \\
\hline
AP Status Check & Verifies whether the AP is active or inactive \\
\hline
AP Connection Count & Checks the number of devices connected to the AP \\
\hline
Connectivity Check & Verifies connectivity with registered devices \\
\hline
\end{tabular}
\label{tab:sdn-functions}
\end{table}

It also supports editing, removing, or querying registered devices, as well as viewing their locations directly on a map or via Google Maps.

\begin{figure}
\centerline{\includegraphics[width=5.5in]{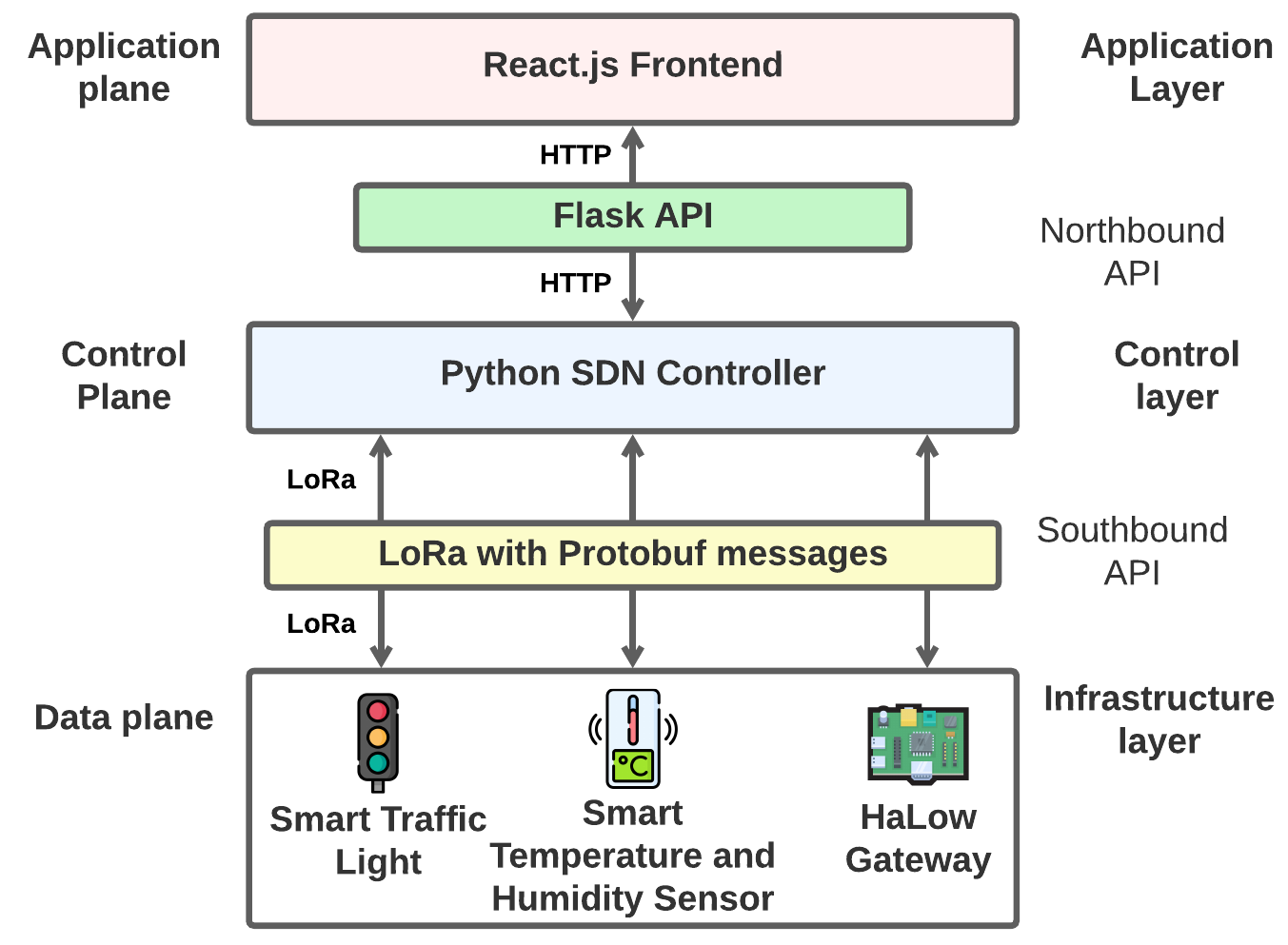}}
\caption{Diagram of the SDN platform architecture, illustrating its planes, layers, and API.
\label{fig:sdn_plataform}}
\end{figure}

Considering the architectural pattern illustrated in \autoref{fig:sdn_plataform}, each device within the control plane communicates with the SDN controller through a LoRa-based controlled flooding mesh protocol using the LTFH mechanism, which has been implemented via a Python script.

The SDN controller communicates with the devices to be configured (Southbound Application Programming Interface (API)) via LoRa and with the SDN web application (Northbound API) via Hypertext Transfer Protocol (HTTP).

\begin{figure}
\centerline{\includegraphics[width=\linewidth]{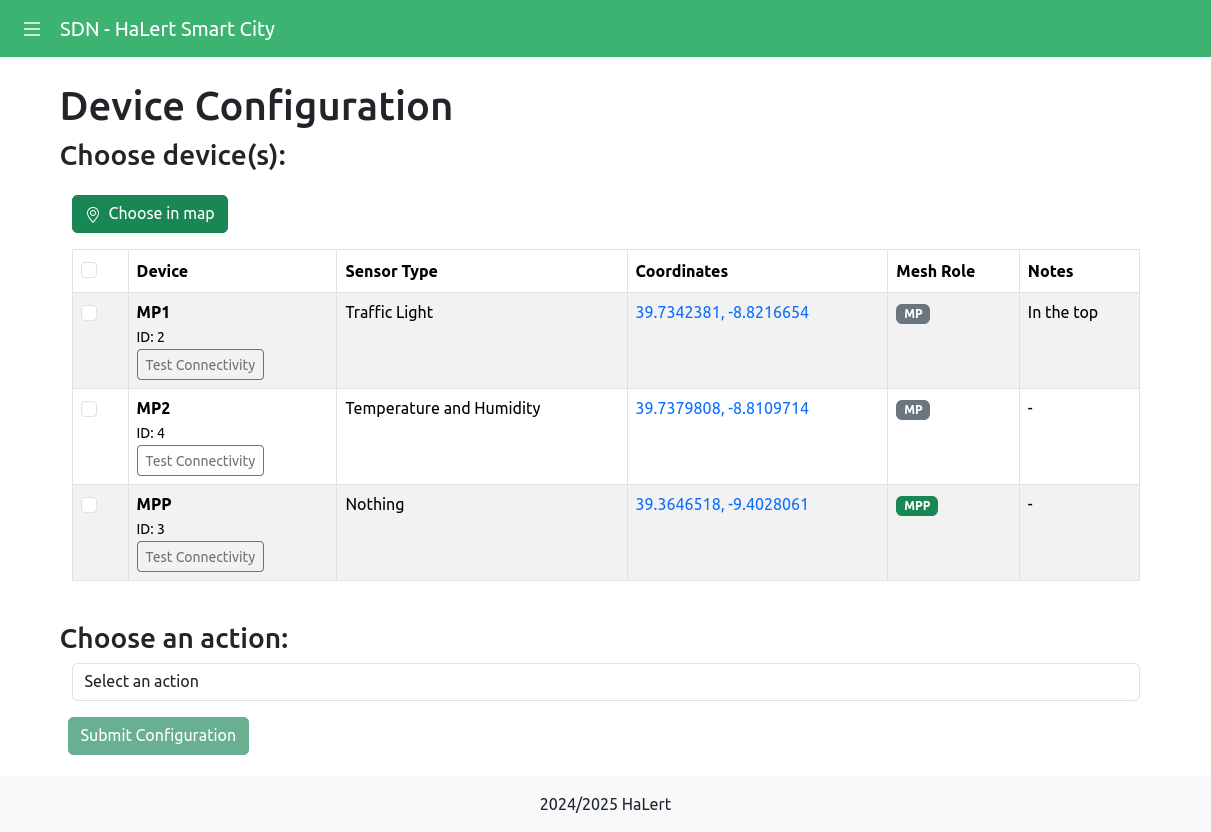}}
\caption{SDN Platform Interface displaying the device configuration form, which includes a device selection table and a choose action field. \label{fig:screenshot_sdn_platform}}
\end{figure}

The web application, hosted locally (internet independent), shown in \autoref{fig:screenshot_sdn_platform} developed using the React.js framework and operating within the application plane, allows configurations to be sent to individual devices, selected groups of devices, or broadcast to the entire network.

In line with our approach described in \citep{Ortigoso2024}, Protobuf messages are employed within the LoRa network. However, an optimisation tailored specifically for this scenario has been introduced, involving the encoding of these messages as dictionaries rather than strings. Consequently, when a user interacts with a controller endpoint via the frontend, a Protobuf message consisting of four \verb|uint32| fields, namely, the source device ID, destination device ID, message ID, and action ID. Encoding these fields as integer values significantly reduces the payload size, achieving an approximate 50\% reduction from the original 80–100 bytes to about 40–50 bytes.

\begin{figure}
    \centering
    \includegraphics[width=4.5in]{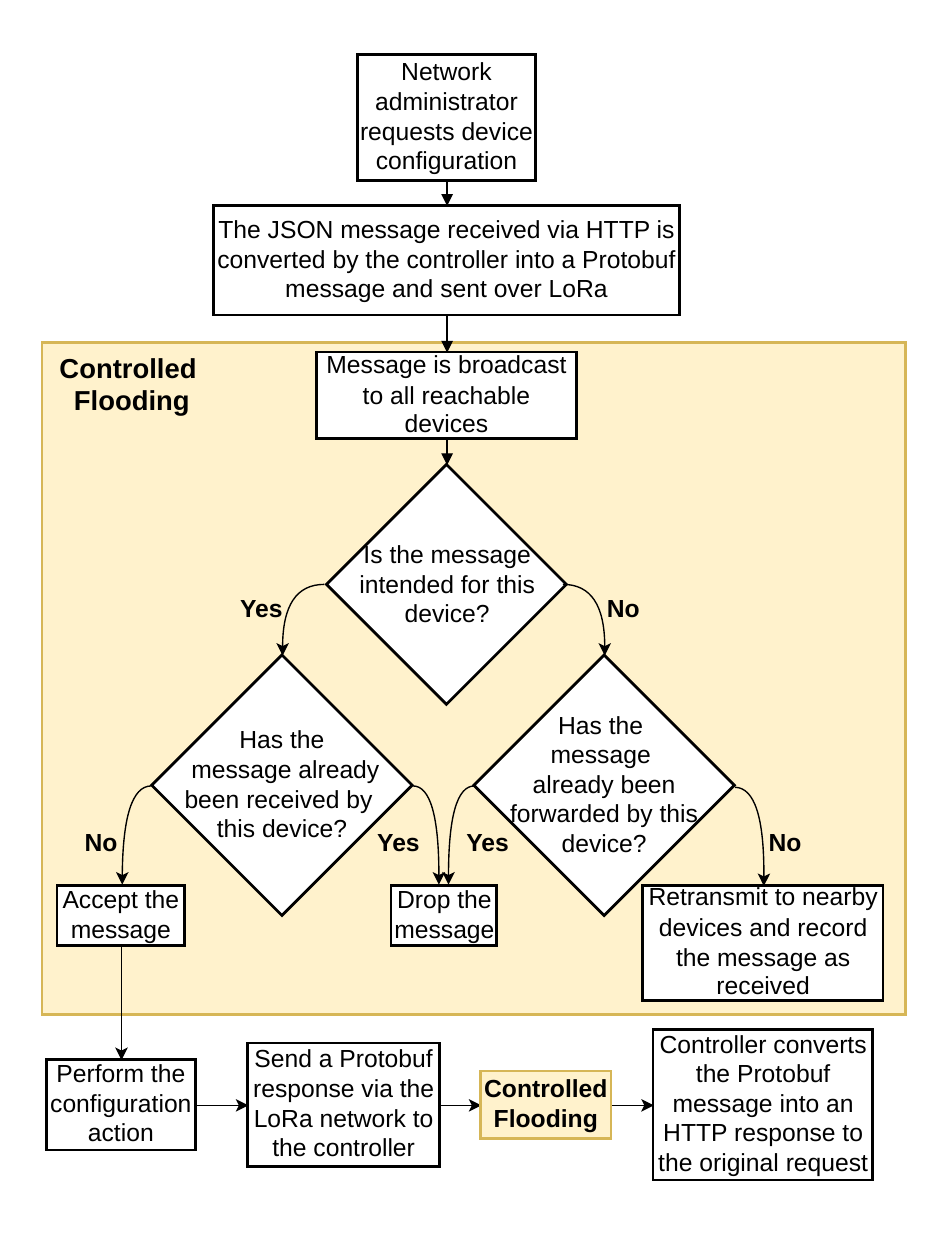}
    \caption{Process of sending configuration messages over LoRa network using controlled flooding.}
    \label{fig:LoRa_SDN_flooding}
\end{figure}

The process of sending a configuration message over the LoRa network with controlled flooding, decipted in \autoref{fig:LoRa_SDN_flooding}, involves the following steps: Initially, a JSON message received via HTTP by the SDN controller is converted into a Protobuf message. This Protobuf message is then broadcasted to all reachable devices. Each device evaluates whether the message is intended for itself and whether it has previously received or forwarded it. If the message is not relevant or already processed, it is dropped. Otherwise, the device accepts the message, executes the configuration action, retransmits it to nearby devices, and records it as received. Subsequently, a Protobuf response message is sent back through the LoRa network to the controller, completing the loop. The controller finally converts this response back into an HTTP message, responding to the initial request.

The process of sending a configuration message over the LoRa network using controlled flooding, depicted in \autoref{fig:LoRa_SDN_flooding}, begins when the network administrator uses the SDN application to send a configuration request to the SDN controller as a JSON message via HTTP. This JSON message is converted into a Protobuf format by the controller and broadcasted to all reachable devices in the network. Each device receiving the broadcast first determines whether the message is specifically intended for it. If not intended, the device checks whether it has previously forwarded this message. If the message has not been forwarded, it is dropped; if it has been forwarded, the device retransmits it to other nearby devices. Conversely, if the message is intended for the device, it verifies whether it has already been received. If already received, the message is dropped; otherwise, the device accepts the message, executes the configuration action, and sends a response back to the controller using the LoRa Network using the same controlled flooding method. Finally, the controller converts the received Protobuf response back into an HTTP response, completing the configuration cycle.

In critical scenarios requiring the rapid and efficient reconfiguration of a large number of devices, payload reduction becomes essential, as it not only enables faster transmission of configuration data but also allows a greater volume of information to be transmitted within a given time frame. This optimisation further reduces radio spectrum occupancy, thereby decreasing interference with other LoRa transmissions and coexisting communications operating on the same frequencies. Such improvements are particularly advantageous, as they minimise the likelihood of message loss or reception errors.

In cases where the response must convey a numeric quantity—such as the number of devices currently connected—a compact encoding scheme is employed. Specifically, the response ID is calculated by multiplying the original action ID by the maximum expected quantity range (minus one), and subsequently adding the actual reported value. Assuming a maximum of 100 possible values, the encoding follows the formula \verb|responseID = actionID × 100 + n|, where \verb|n < 100|. For instance, if the action ID associated with querying the number of connected devices is 5, and the system reports that 3 devices are currently connected, the response ID becomes \verb|5 × 100 + 3 = 503|. This method enables the controller to infer both the type of response and its corresponding value directly from the numeric response ID, without the need for additional payload data.

%% file: Sections/4DevelopedPrototype/IoTPlatform.tex
\subsection{IOT PLATFORM}

The IoT platform is responsible for managing and monitoring the sensing and actuation activities within the city. Each IoT device deployed across the urban environment runs a Python script to periodically transmit status data to an MQTT broker hosted on the HaLow gateway. The broker was configured using Mosquitto, with both standard MQTT communication and MQTT over WebSockets enabled to facilitate seamless integration with the web application.

\begin{figure}
\centerline{\includegraphics[width=5.5in]{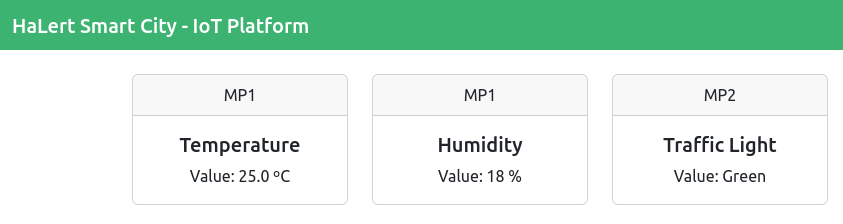}}
\caption{IoT platform interface displaying the detected sensor values, indicating both the device on which each sensor is installed and the corresponding sensor name.\label{fig:screenshot_iot_plataform}}
\end{figure}

Furthermore, the developed web application in React.js, shown in \autoref{fig:screenshot_iot_plataform}, displays the data published to the MQTT broker. This enables city management personnel to monitor the real-time status of the smart city infrastructure.

%% file: Sections/4DevelopedPrototype/EmergencyChats.tex
\subsection{EMERGENCY CHATS}

\begin{figure}[]
    \centering
    \subfloat[Citizens access in captive portal]{%
        \includegraphics[width=2.4in]{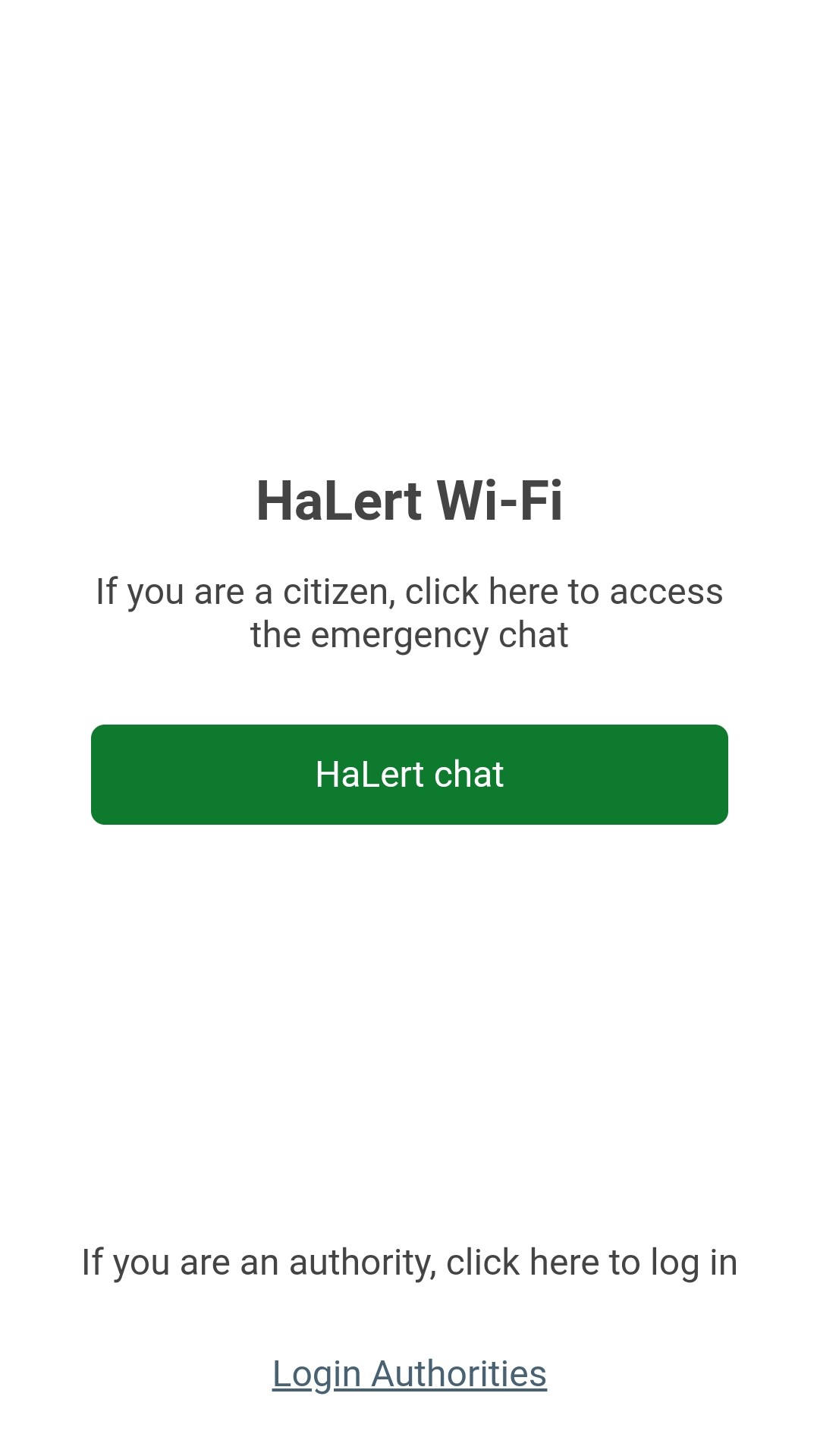}
    }
    \hspace{0.2in}
    \subfloat[Authority login in captive portal]{%
        \includegraphics[width=2.4in]{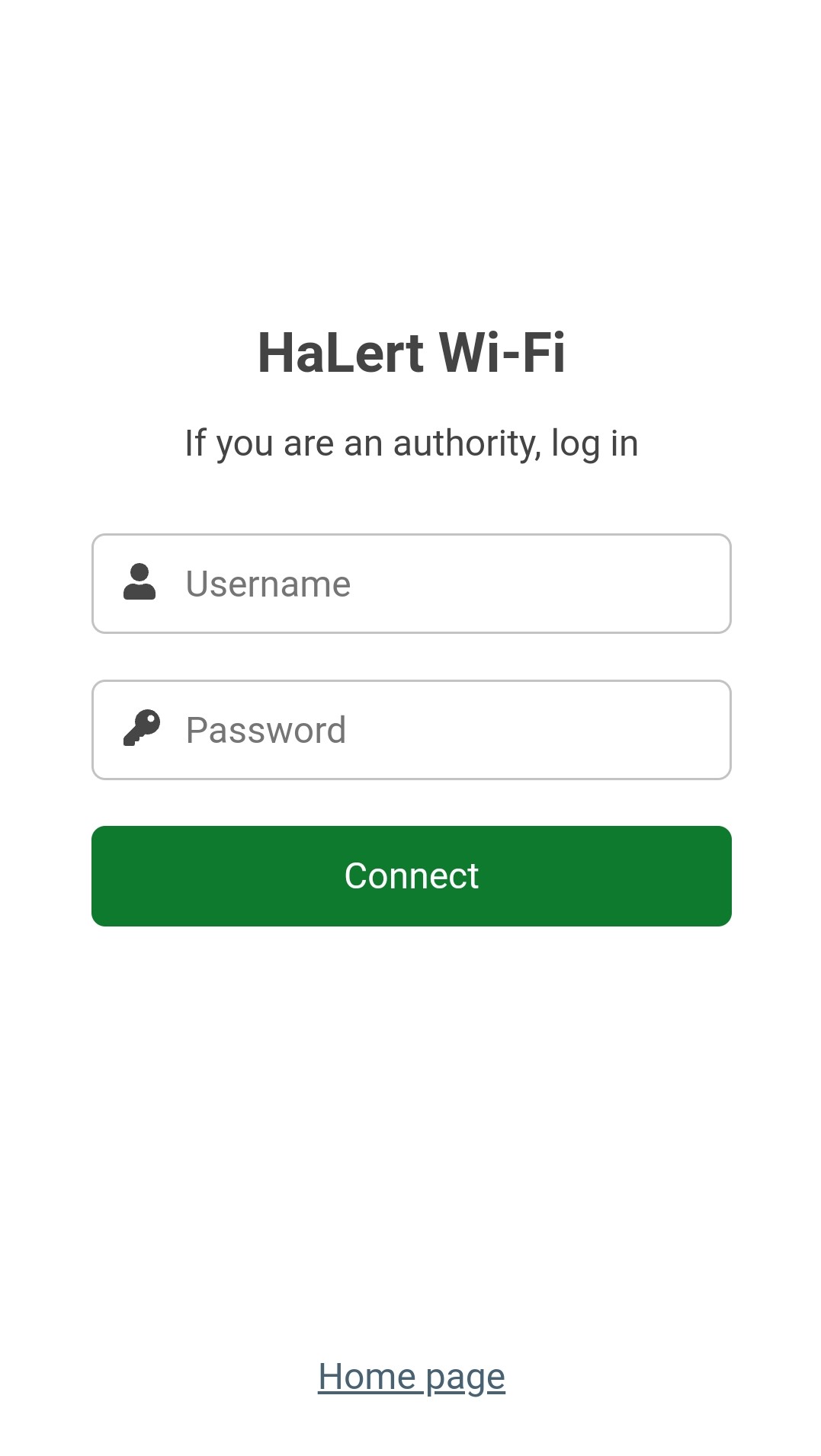}
    }
    \caption{Access point captive portal. (a) Page displaying a button for citizens access to the citizen chat. (b) Page containing a login form for authorities to access restricted pages, including the authority chat.}
    \label{fig:screenshots_captive_portal}
\end{figure}

In emergency scenarios, Wi-Fi access points are made available to the public and authorities to facilitate access to dedicated web-based emergency chat applications. These access points, provided by Mikrotik mAP lite routers, offer captive portals (referred to as ``hotspots'' by Mikrotik), shown in \autoref{fig:screenshots_captive_portal}, through which citizens can access the emergency chat designated for public use. Authorities, upon authentication, can access not only the public emergency chat but also a restricted emergency communication channel intended exclusively for authorities, along with internet access, if available.

When accessing either chat service, users (citizens or authorities) are required to enter a display name by which they will be identified within the chat.

\begin{figure*}
    \centering
    \subfloat[Citizens chat]{%
        \includegraphics[width=2.4in]{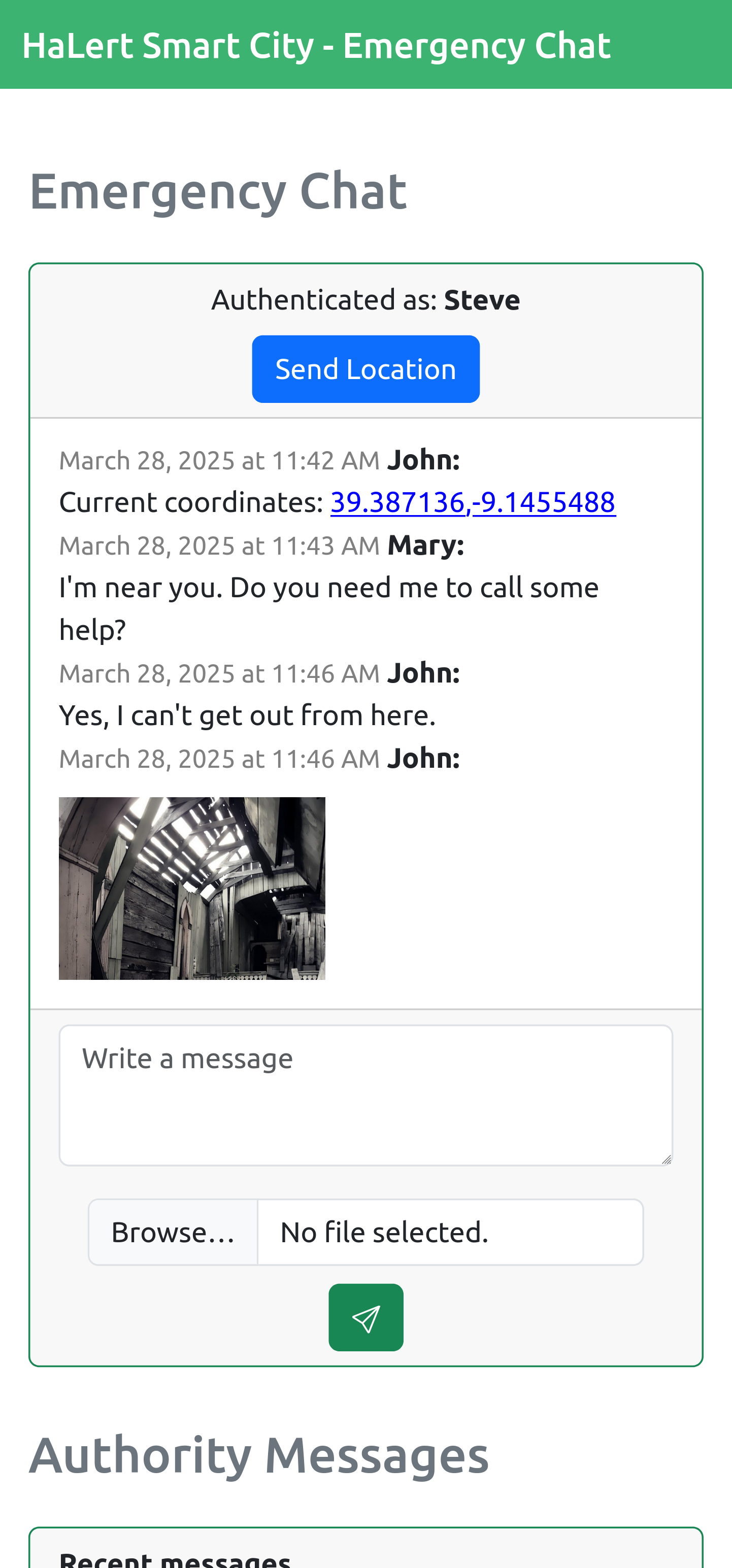}
    }
    \hspace{0.2in}
    \subfloat[Authority messages]{%
        \includegraphics[width=2.4in]{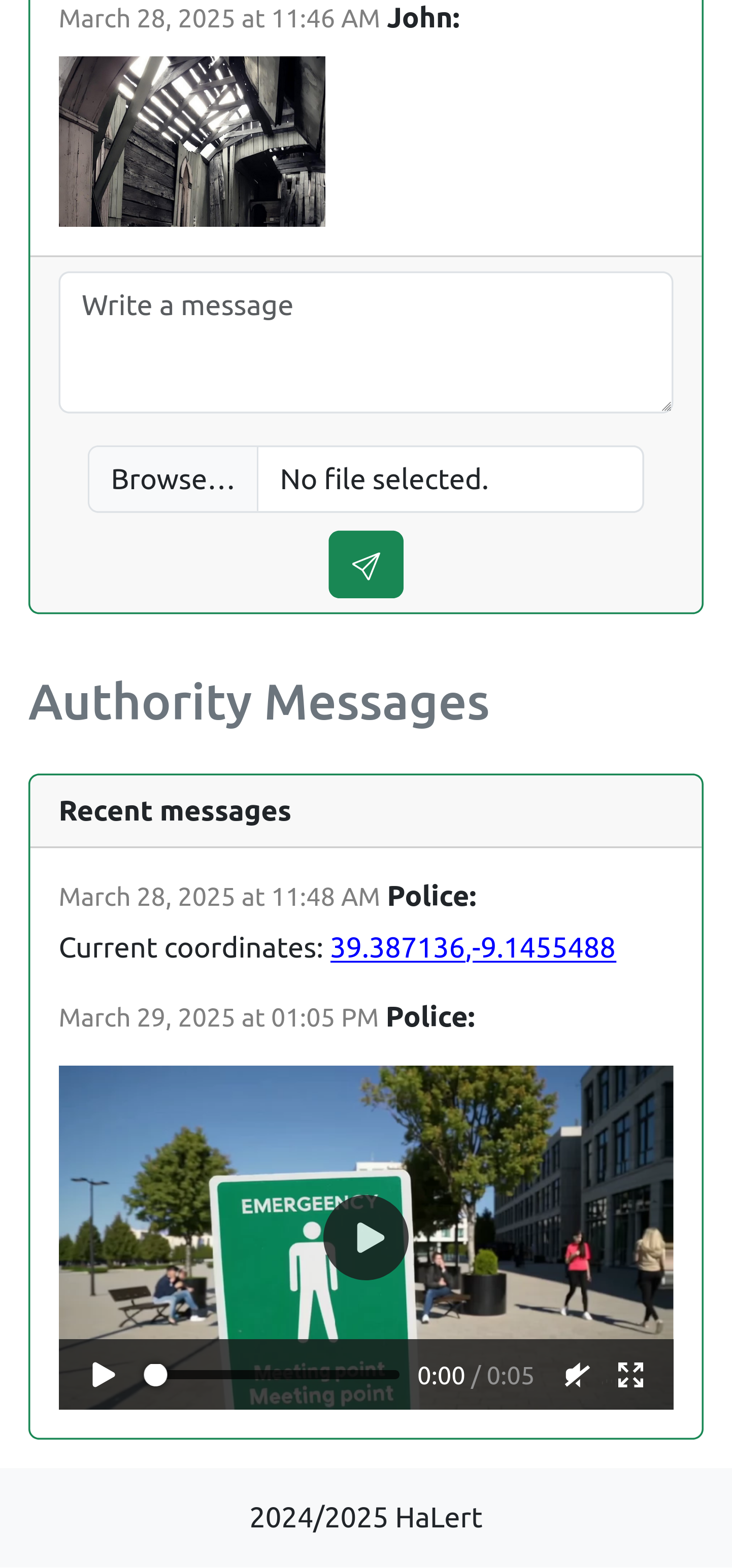}
    }
    \caption{Citizen emergency chat interface. In the citizen message area (a), users can send and view text, location, and image messages. In the authority message area (b), users can view text, location, image, audio, and video messages.}
    \label{fig:citizens_emergency chat}
\end{figure*}

As displayed in \autoref{fig:citizens_emergency chat}, in the citizen emergency chat users can exchange text messages, share their current coordinates and upload images. In addition, citizens have access to alert messages published by the authorities.

\begin{figure}
\centerline{\includegraphics[width=2.5in]{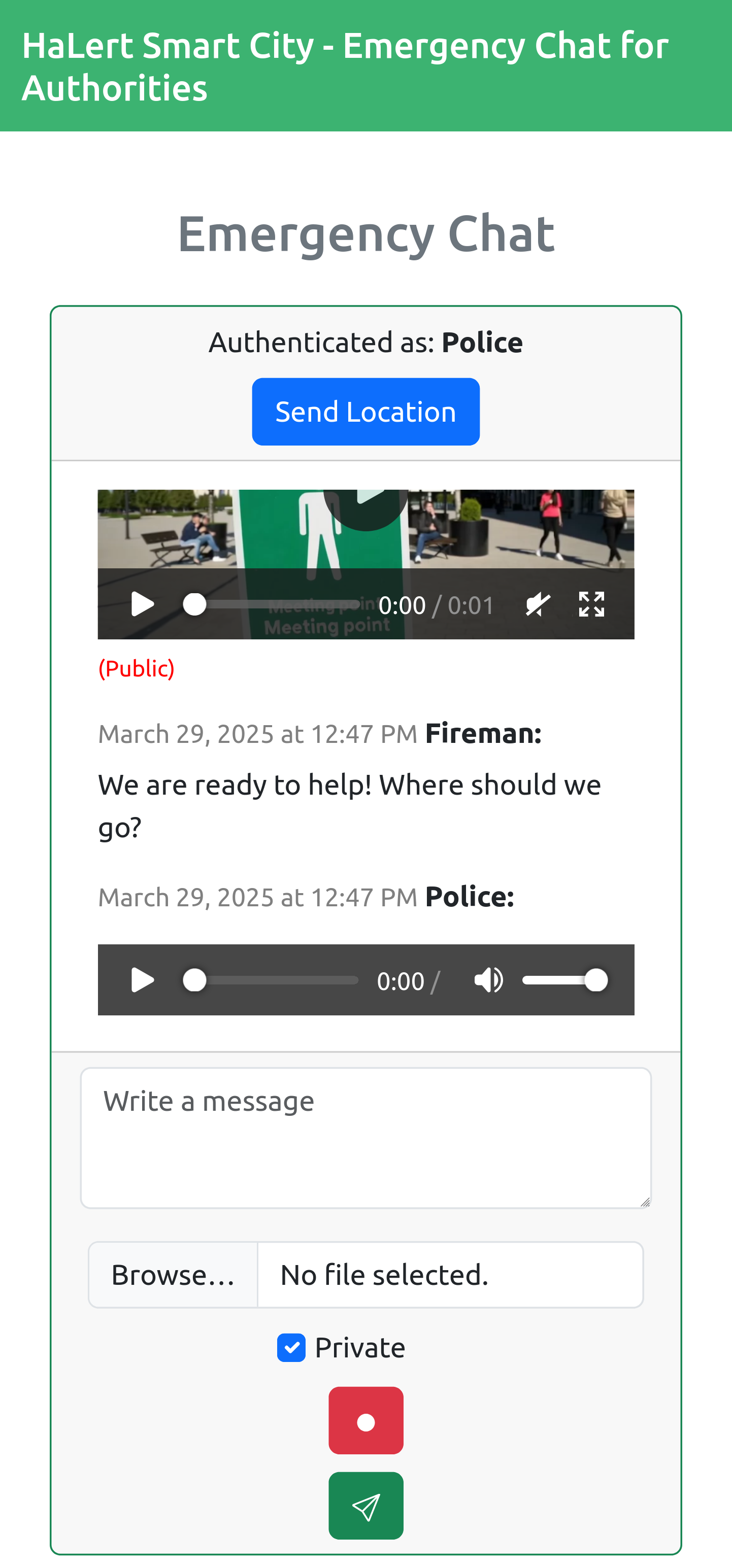}}
\caption{Authorities emergency chat, where authorities can exchange text, location, image, audio, and video messages.\label{fig:screenshot_authorities_chat}}
\end{figure}

In the authorities emergency chat, authorised users can exchange text messages, coordinates, videos, voice messages, and images amongst themselves. They also have the ability to send alert messages of these types to all citizens by specifying whether the message is public or private, as illustrated in \autoref{fig:screenshot_authorities_chat}.

The decision to restrict the types of media messages that citizens can send is based on the significantly larger number of citizen users compared to authorities and their lack of specialised training in communication protocols. Authorities are trained to adhere to specific guidelines, such as sending videos of minimal necessary length and structured clearly for effective interpretation by other users. Furthermore, citizens are intentionally denied internet access, even when it is technically available, as unrestricted internet access in emergency situations is non-essential and may overload the system, potentially leading to service denial and hindering the transmission of urgent and critical communications.

Both emergency chat applications operate using a Node.js backend that employs WebSockets for real-time message transmission. The backend also provides specific endpoints for retrieving message history and serving file content. As the current system is a prototype, the historical data of the latest 5000 messages is stored in a JavaScript Object Notation (JSON) file, facilitating straightforward migration to a non-relational database in production scenarios.

The frontend interfaces of these applications were developed using the React.js framework and implemented as Progressive Web Apps (PWA), allowing users to access them directly via a browser or install them locally. When installed, PWA reduce network usage by caching resources, resulting in improved performance and offline accessibility.

Several mechanisms were implemented to enhance the efficiency and responsiveness of the emergency chat applications, particularly in resource-constrained environments common during disaster scenarios. Image and video files are compressed on the frontend before being transmitted to the backend, significantly reducing payload size and upload time. This not only decreases bandwidth usage but also minimises congestion on the network, allowing more users to exchange critical information simultaneously. The adoption of the OGG format for audio messages ensures efficient encoding with reduced file sizes without compromising intelligibility, which is particularly important when transmitting over networks with limited capacity. On the backend, additional compression techniques are applied to static assets and dynamic content, improving page load times and reducing server-side response delays. Furthermore, to maintain application performance and avoid excessive memory consumption, the system restricts the retrieval of stored messages to the most recent 500 entries. This approach ensures quick access to relevant conversation history while preventing the accumulation of large datasets that could degrade the user experience or overload devices with limited processing power. Collectively, these optimisation strategies improve message delivery latency and maintain the stability and usability of the platform under adverse network conditions.

%% file: Sections/5Tests.tex
\section{TESTS AND RESULTS}

\begin{figure*}
\centerline{\includegraphics[width=\linewidth]{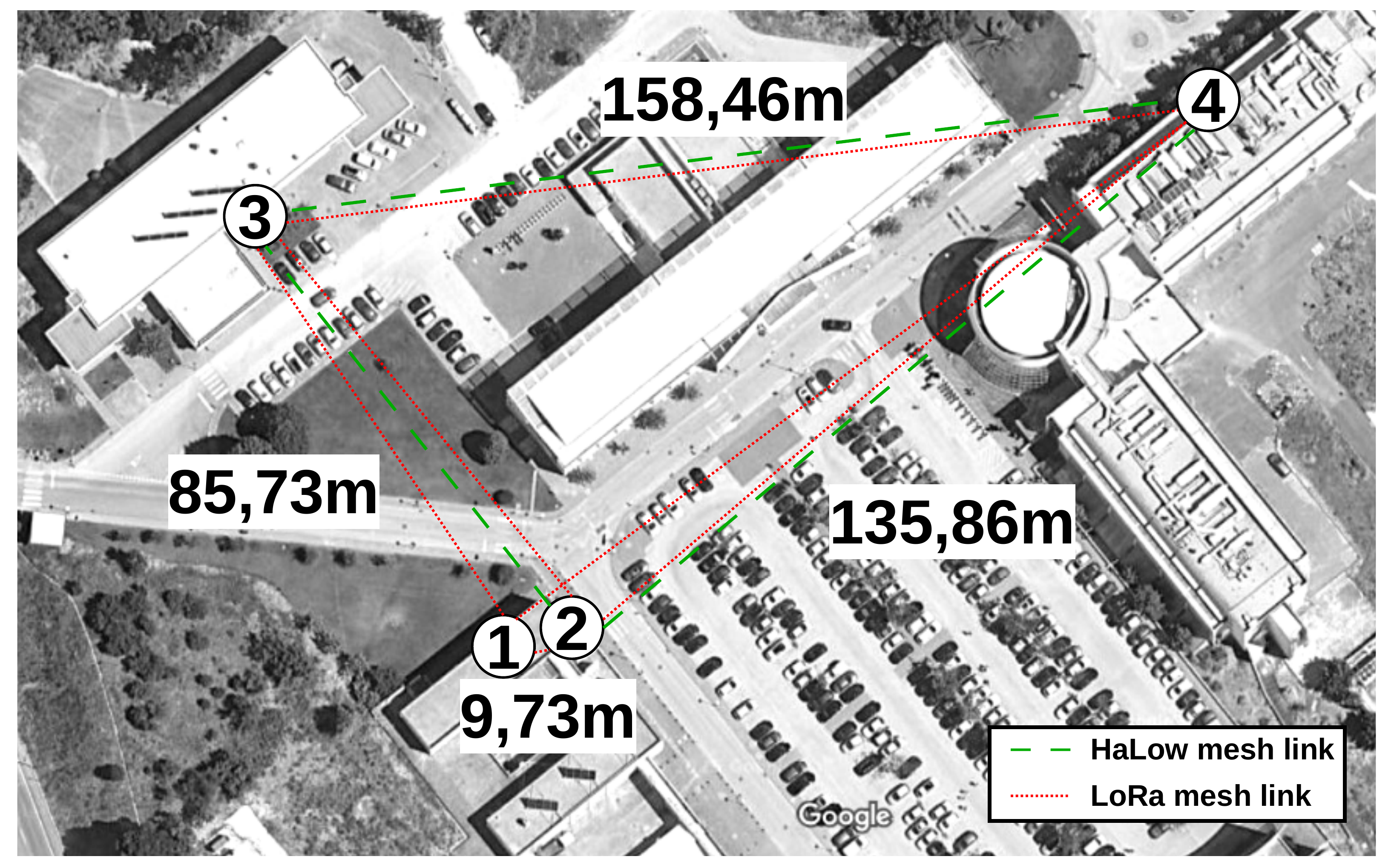}}
\caption{Campus device locations during testing at Campus 2 of the Polytechnic University of Leiria. (1) represents the location of the LoRa gateway, (2) the HaLow gateway, (3) the smart temperature and humidity sensor, and (4) the smart traffic light.\label{fig:test_map}}
\end{figure*}

\begin{figure}[]
    \centering
    \subfloat[HaLow gateway]{%
        \includegraphics[width=0.48\linewidth]{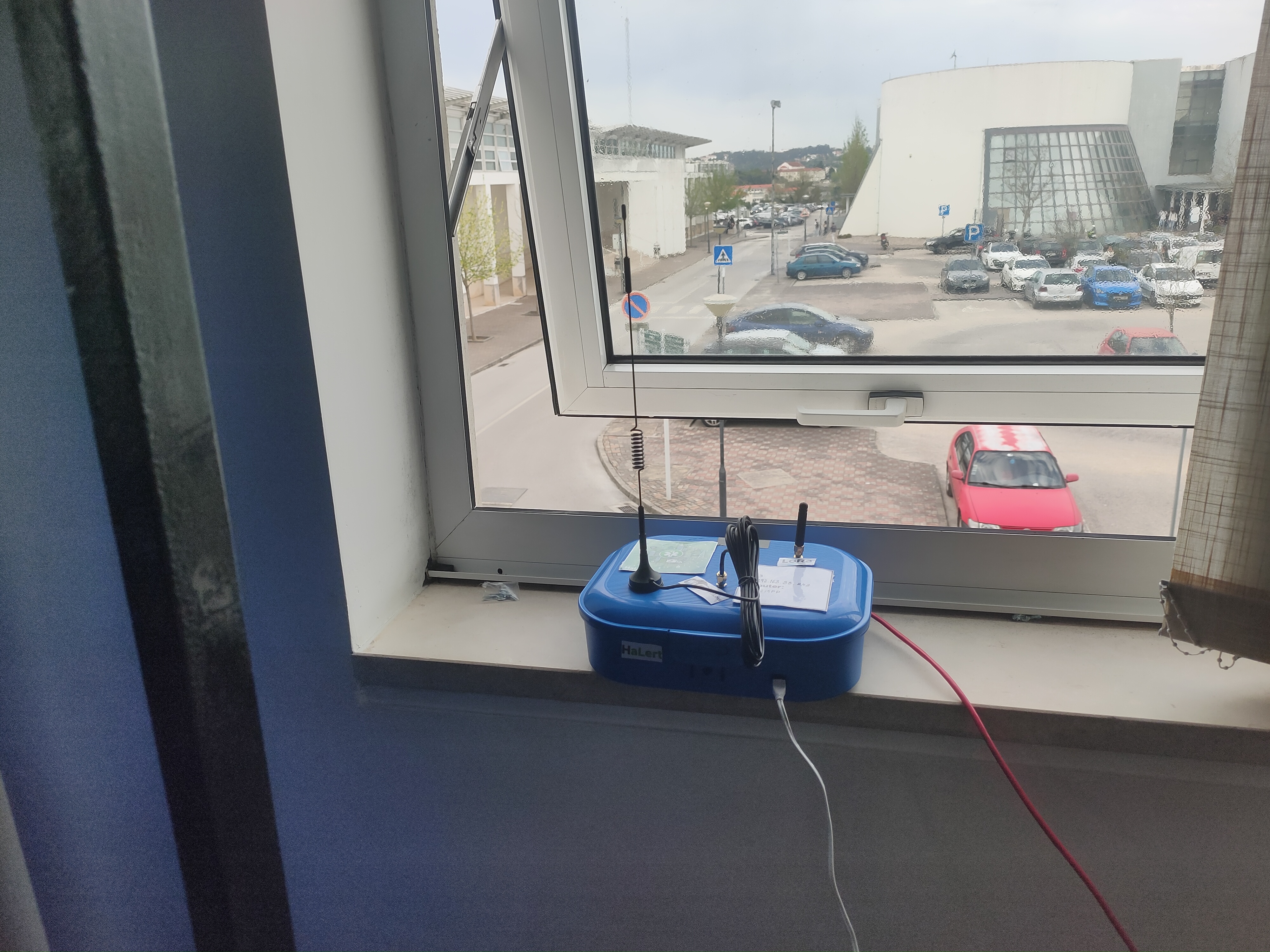}
        \label{subfig:image_halow_gateway}
    }
    \hfill
    \subfloat[Smart temperature and humidity sensor]{%
        \includegraphics[width=0.48\linewidth]{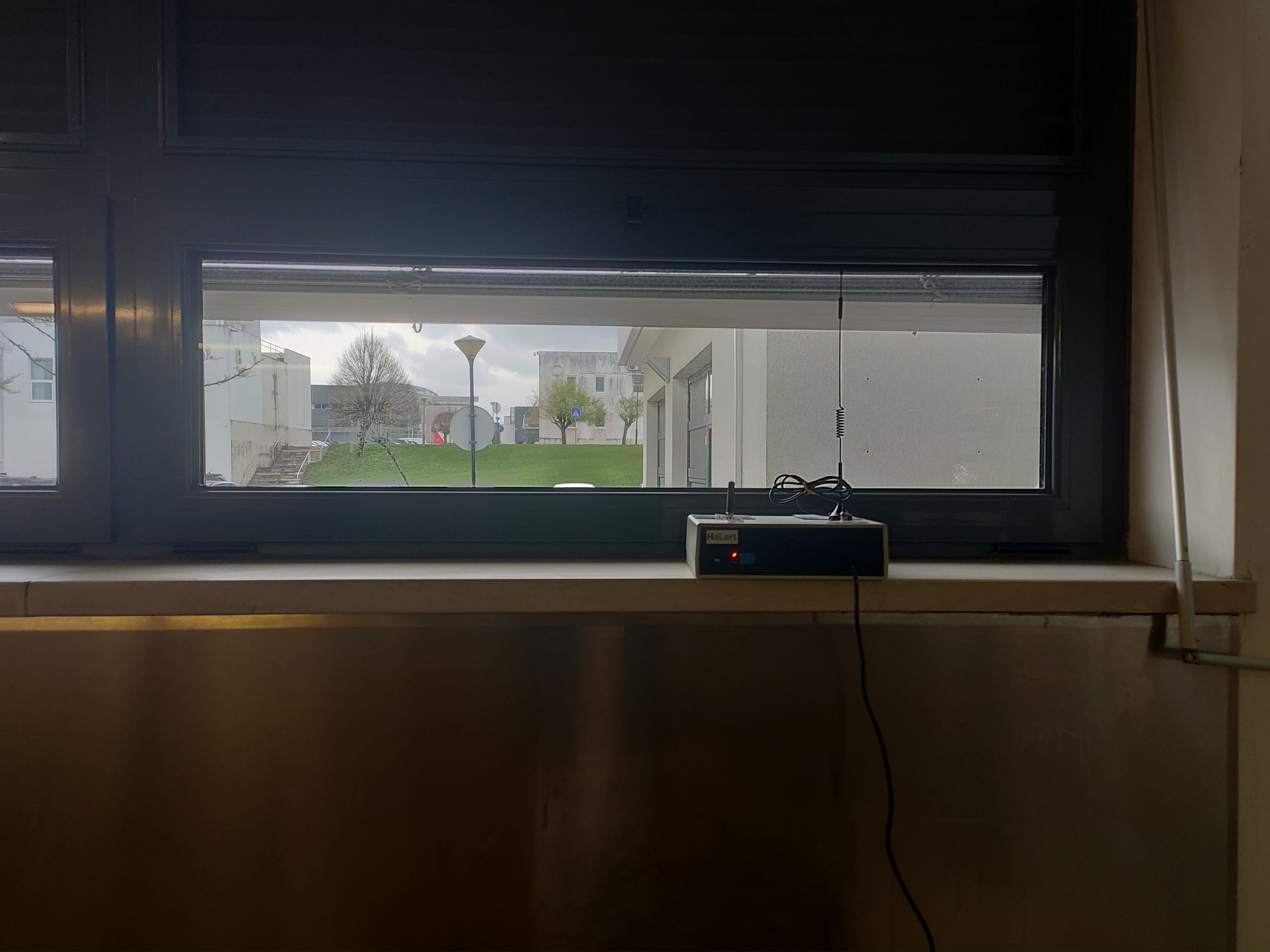}
        \label{subfig:image_smart_temperature_humidity_sensor}
    }
    \hfill
    \subfloat[Smart traffic light]{%
        \includegraphics[width=0.60\linewidth]{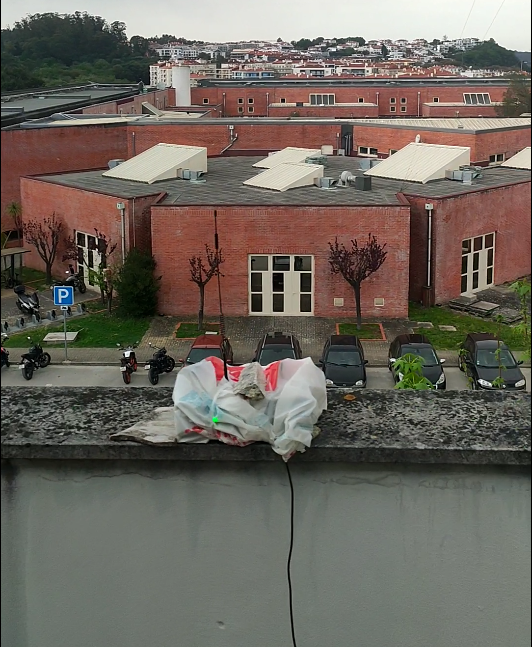}
        \label{subfig:image_smart_traffic_light}
    }
    \caption{Photos of device locations during testing. (a) The HaLow gateway positioned behind an open window. (b) The smart temperature and humidity sensor placed behind a closed window. (c) The smart traffic light situated inside a bag on a balcony ledge.}
    \label{fig:images_test_location}
\end{figure}

To evaluate the developed prototype, practical tests were conducted at Campus 2 of the Polytechnic University of Leiria. The equipment was strategically placed as illustrated on the map in \autoref{fig:test_map}. Specifically, the LoRa gateway (1) was installed inside a room located on the first floor of the building, while the HaLow gateway (2) was placed behind an open window of the same room, as illustrated in Figure \ref{subfig:image_halow_gateway}. The Smart Traffic Light (4) was positioned on the balcony ledge on the second floor of the building, as shown in Figure \ref{subfig:image_smart_traffic_light}. Finally, the Smart Temperature and Humidity Sensor (3) was located on the ground floor of another building,  in close proximity to a closed window, as illustrated in Figure \ref{subfig:image_smart_temperature_humidity_sensor}. It is important to note that the ground floor of the building housing the Smart Temperature and Humidity Sensor is located approximately four metres below the building where the HaLow gateway is installed.

These locations and scenarios were specifically selected to realistically assess the performance and functionality of the proposed solution in an urban environment characterised by dense infrastructure, various obstacles, and varying terrain elevations. The distances indicated in \autoref{fig:test_map} were measured horizontally, in straight lines, without accounting for altitude differences between devices. Consequently, the actual distances are greater than those represented in the figure.

Two distinct types of tests were conducted, each with specific objectives. The first aimed to evaluate the performance parameters of the HaLow and LoRa networks, focusing on metrics such as signal strength, latency, throughput, and message success rate, in order to assess the communication reliability and stability of each network under realistic urban conditions. The second set of tests was functional in nature and intended to validate the practical operation of the prototype. These tests verified whether all implemented features were operational and effective within the deployed network environment.

The evaluation of the HaLow network involved conducting all tests from the MPP (HaLow gateway), accessed via Secure Shell (SSH), towards two MP (smart devices). Initially, Received Signal Strength Indicator (RSSI) and Signal-to-Noise Ratio (SNR) values were assessed using the command \verb|show signal| from the \verb|cli_app| provided within the \verb|nrc7292_sw_pkg| software package. For the smart temperature and humidity sensor, the RSSI measured was -91 with an SNR of 2, whereas the smart traffic light presented an RSSI of -84 and an SNR of 15.

\begin{figure}
\centerline{\includegraphics[width=4.5in]{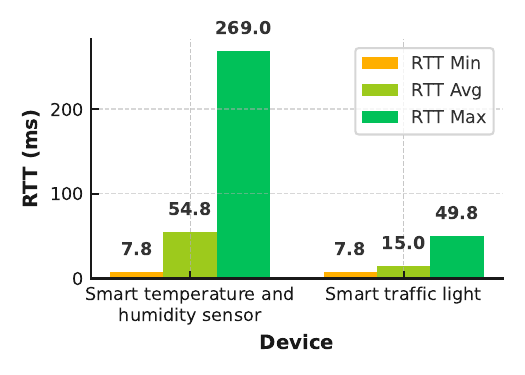}}
\caption{RTT minimum, average and maximum from HaLow gateway to smart devices\label{fig:latency_chart}}
\end{figure}

Subsequently, latency and throughput assessments were conducted. Latency evaluation employed the \verb|ping| utility, involving three sets of 20 ICMP packets, each with a size of 64 bytes. It was observed that throughout all tests, packet loss was consistently 0\%. As demonstrated in \autoref{fig:latency_chart}, which displays the minimum, maximum, and average Round-Trip Times (RTT) in milliseconds for the three attempts per smart device, the average and maximum RTT were considerably higher for the smart temperature and humidity sensor compared to the smart traffic light (54.8 ms vs 15 ms for average RTT; 269 ms vs 49.8 ms for maximum RTT). However, the minimum RTT value was identical for both smart devices, at 7.8 ms.

\begin{figure}
\centerline{\includegraphics[width=4.5in]{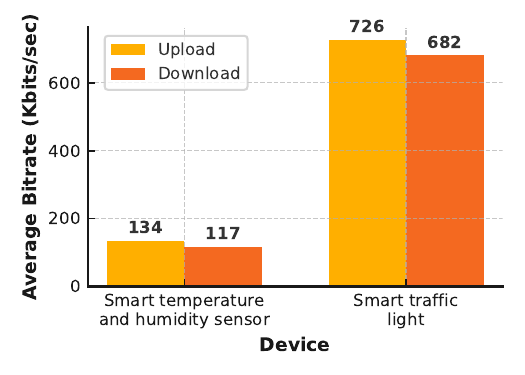}}
\caption{Average upload and download bitrate from HaLow gateway to smart devices\label{fig:bitrate_chart}}
\end{figure}

A similar behaviour was noted during the bitrate evaluation, with results detailed in \autoref{fig:bitrate_chart}. The results indicate substantially higher upload and download bitrates for the smart traffic light compared to the smart temperature and humidity sensor (upload bitrate: 726 Kbits/sec vs 134 Kbits/sec; download bitrate: 682 Kbits/sec vs 117 Kbits/sec). This evaluation utilised the \verb|iPerf| utility, with the server operating on the MPP and the clients accessed and operated through a Python script executing SSH commands on the smart devices. Each client was subjected to three tests, each lasting 60 seconds. The evaluation methodology was based on the approach described in \citep{Kane2023}.

These findings highlight the significant impact of obstacles on Wi-Fi HaLow transmissions, even over relatively short distances, and the consequent effect on both bitrate and latency performance. Nevertheless, network stability was relatively unaffected, as evidenced by the absence of packet loss. Furthermore, these results underscore Wi-Fi HaLow's potential effectiveness in mixed medium-range urban scenarios (both indoor and outdoor).

To evaluate the performance of the LoRa network, a Python script was developed, which sequentially sends requests to the REpresentational State Transfer (REST) API of the SDN controller, awaits responses, and pauses for one second before initiating the next request. This script was executed continuously for a period of one hour, during which data was collected from each LoRa-connected device (smart temperature and humidity sensor, smart traffic light, HaLow gateway and LoRa gateway). The collected data comprised total counts of error messages, retransmitted messages, received messages, sent messages, and ignored messages.

Error messages are defined as messages that encountered reception errors. Within the total successfully received messages: retransmitted messages refer to those received by a device but intended for forwarding to another device; received messages are those directly intended for the receiving device itself; and ignored messages are duplicates already received by the same device.

\begin{figure}
\centerline{\includegraphics[width=4.5in]{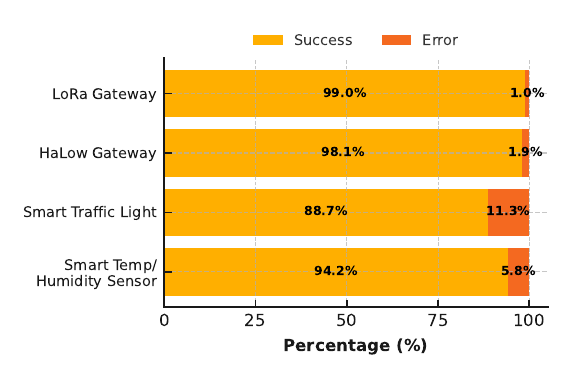}}
\caption{Success and error rates in LoRa received messages\label{fig:lora_messages_rate}}
\end{figure}

The analysis of this data, illustrated in \autoref{fig:lora_messages_rate}, indicates that the success rate of the received messages ranged from 88.7\% to 99\%, averaging 94.96\%. Conversely, the error rate of received messages varied between 1.03\% and 11.35\%, with an average of 5.04\%.

\begin{figure*}
\centerline{\includegraphics[width=\textwidth]{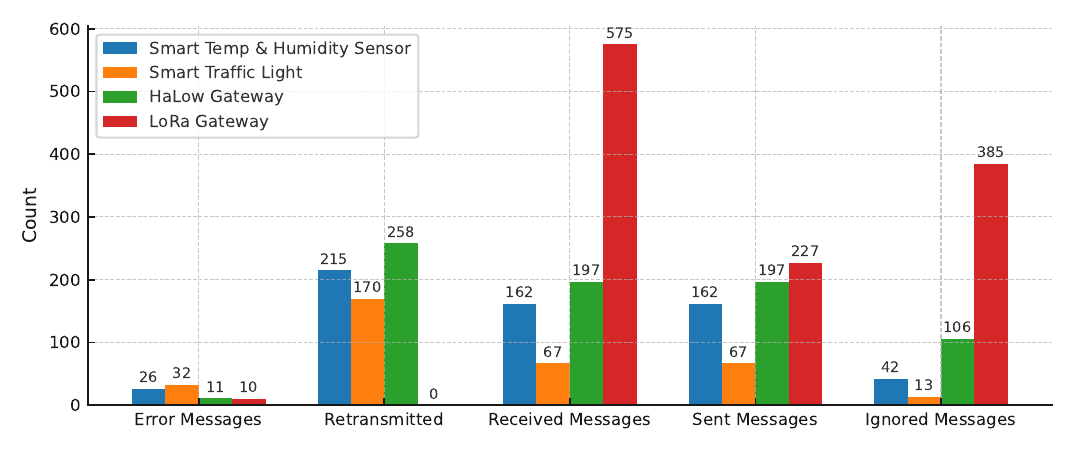}}
\caption{Detailed statistics of LoRa messaging using LTFH and a controlled flooding mesh during testing across all devices. The data includes the number of error messages, retransmitted messages, received messages, sent messages, and ignored messages.}
\label{fig:lora_messages_chart}
\end{figure*}

Observing the detailed results from the LoRa network tests presented in \autoref{fig:lora_messages_chart}, it is evident that the number of messages received with errors by smart devices (26 for the smart temperature and humidity sensor and 32 for the smart traffic light) is approximately three times higher than that received by the gateways (11 for the HaLow gateway and 10 for the LoRa gateway). While the precise cause of this discrepancy remains unconfirmed, it is reasonable to hypothesise that it is attributable to the physical locations of the devices. Devices positioned at greater distances and facing more physical obstructions relative to the originating LoRa gateway are likely to experience increased reception errors.

Regarding retransmitted messages, the LoRa gateway notably did not retransmit any messages. This behaviour can be logically explained by the fact that the LoRa gateway only device on the network that originates messages (excluding responses to commands); thus, retransmission of its own messages is unnecessary. For other devices, the number of retransmitted messages ranged between 170 and 258 messages, displaying limited variability.

Looking at the number of received messages, as expected, the LoRa gateway recorded the highest number, reflecting its role as the primary recipient of responses following the execution of commands or scripts on other devices. Additionally, the number of messages received and sent by devices other than the LoRa gateway is equal, as each received configuration message elicits a response.

Finally, the distribution of ignored messages closely mirrors that of received messages, an anticipated outcome since ignored messages are also specifically directed to individual devices.

The functional tests were designed to evaluate the system’s operational capabilities under the defined network conditions. Specifically, they examined whether it was possible to remotely reconfigure the devices, thereby verifying the effectiveness of remote management functionalities. Additionally, the tests assessed whether real-time monitoring of the sensors status could be reliably conducted through the web-based dashboard, ensuring proper data transmission and interface responsiveness.  

Further aspects included verifying access to the access points (AP) and their associated captive portals using smart devices, an essential feature for user onboarding and network interaction. The tests also examined whether users could access the citizen chat interface without prior authentication, as well as whether, upon successful login, access to the authority chat and broader internet services was enabled. The functionalities evaluated and the corresponding results are summarised in \autoref{tab:functional_validation}.

\begin{table}
\caption{Summary of functional tests, detailing the functionalities evaluated and the corresponding results obtained. Device numbering follows the convention used in \autoref{fig:test_map}.}
\centering
\begin{tabular}{|p{6cm}|p{9.5cm}|}
\hline
\textbf{Functionality Tested} & \textbf{Result} \\
\hline
Remote reconfiguration of devices & Successful remote reconfiguration of devices (2), (3), and (4) via the SDN platform \\
\hline
Real-time sensor monitoring & Correct and timely visualisation of sensor data from devices (3) and (4) on the web dashboard \\
\hline
Access to Wi-Fi APs via smart devices & Functional Wi-Fi access confirmed on devices (3) and (4) \\
\hline
Captive portal accessibility & Captive portals displayed correctly on devices (3) and (4) \\
\hline
Unauthenticated access to citizen chat & Citizen chat accessible without login through captive portal on devices (3) and (4) \\
\hline
Authenticated access to authority chat and internet & Successful login and full access to authority chat and internet on devices (3) and (4) \\
\hline
\end{tabular}
\label{tab:functional_validation}
\end{table}

As shown in \autoref{tab:functional_validation}, all functionalities were thoroughly tested and were confirmed to be fully operational within the constraints of the established network environment, thereby demonstrating the robustness and reliability of the prototype in realistic usage scenarios.

%% file: Sections/6Discussion.tex
\section{DISCUSSION}

The experimental results confirm the feasibility of the proposed architecture in supporting both routine IoT data exchange and the dynamic reallocation of computational resources during emergency scenarios. The developed prototype was not a full-scale smart city deployment, but rather a functional validation of the proposed architecture under constrained, yet realistic, mixed urban conditions. Although the prototype was not implemented in a complete smart city environment, the results suggest that extrapolating the architecture to such a context is both viable and desirable. Nevertheless, this extrapolation must be critically contextualised: smart cities are inherently heterogeneous, comprising complex topologies, variable densities of infrastructure, and fluctuating communication demands. While the prototype environment reproduced some of these conditions (e.g., mixed indoor/outdoor settings, obstacles, elevation differences), further studies are necessary to assess scalability and interoperability at city scale.

An important clarification regarding the network topology is necessary: although the deployed equipment was configured in mesh mode, the evaluation scenario inherently resulted in a star-like topology due to the limited number of available devices. Consequently, extensive mesh functionality testing was not feasible. Addressing this limitation by conducting broader-scale mesh network evaluations represents a important area for future research.

The architecture supports the operationalisation of resilient post-disaster communication systems for citizens and emergency authorities through dual-mesh network deployment. The proposed system leverages Wi-Fi HaLow mesh networking for IoT data transmission and post-disaster communications, while employing a LoRa mesh network to facilitate remote monitoring and configuration via SDN.

However, consistent with previous findings reported in \citep{Thangadorai2025}, the performance and range of the Wi-Fi HaLow mesh network are significantly affected by environmental factors, particularly the lack of direct line-of-sight and physical obstructions. Our experimental scenarios further exacerbate these challenges, combining both indoor and outdoor environments with varying terrain elevations. This is evident from observed performance metrics, specifically the average latency of 54.8 ms, and limited bitrates (upload: 134 Kbps; download: 117 Kbps).

Additionally, hardware and driver instabilities associated with NRC7292-based devices, as previously noted in the literature \citep{Kane2023, Ortigoso2024a}, likely contributed to the observed performance degradation. Our study was also constrained by practical limitations, including a restricted number of test devices and the exclusive use of AHPI7292S and AHMB7292S modules, which may not fully represent the capabilities of other available Wi-Fi HaLow hardware.

In contrast, LoRa performance tests indicated robust message transmission with success and error rates closely aligned with earlier controlled experiments documented in \citep{Ortigoso2024}. With this results we believe that LoRa-based flooding mesh networks can be applied to realistic situations with a satisfactory performance and reliability. Nevertheless, despite these advantages, using LoRa for IoT communication in our use case is not suitable due to inherent limitations in data rate and air time, due to restrictive duty cycle in various locations around the globe. Specifically, the limited data rate precludes simultaneous photo, video and audio transmission and retrieval by multiple users, as well as accessing historical messages. If historical message retrieval were attempted, real-time message reception would be severely compromised due to the low data rate. Conversely, adopting an approach that eliminates message history, thus providing users only with messages sent after joining the chat, would risk missing critical alerts and previous messages, potentially leading to significant public safety concerns.

%% file: Sections/7ConclusionandFutureWork.tex
\section{CONCLUSION AND FUTURE WORK}

This paper proposed, implemented, and evaluated HaLert, a smart city-oriented architecture combining a Wi-Fi HaLow mesh network with a LoRa-based control-plane leveraging controlled flooding. Designed for post-disaster communication, HaLert enables dynamic resource reallocation, supports communication between civilians and authorities, and provides remote monitoring through SDN.

Experimental evaluations in constrained mixed indoor/outdoor urban settings confirmed the system’s feasibility. Despite the Wi-Fi HaLow network exhibiting sensitivity to obstacles, lack of line-of-sight, and elevation changes—affecting latency (15–54.8 ms) and bitrates (134–726 Kbps uplink, 117–682 Kbps downlink)—the network remained stable, with zero packet loss and consistent connectivity, a critical outcome validating its operational reliability.

The LoRa mesh network, employing LTFH and controlled flooding, achieved a 94.96\% message success rate and a 5.04\% error rate, confirming its robustness for configuration and control tasks.

These results validate the functional integration of all system components, highlighting HaLert’s potential to maintain essential communication during infrastructure failure by repurposing urban IoT infrastructure without Internet dependency.

Future research should include simulations of realistic disaster conditions, quantifying service capacity (text, location, image, audio, video) per HaLow access point, and assessing the impact of increased mesh hops on service quality. This will inform the required access point density per 100 users.

Further steps include implementing a prototype integrating HA and load-balancing mechanisms, conducting large-scale scalability tests with more devices and alternative HaLow-compliant hardware, and tackling the limited availability of such hardware in Europe, which remains a barrier to large-scale experimentation.

Lastly, the development of a digital twin or shadow for real-time monitoring and predictive analytics is recommended to enhance system usability, resilience, and maintenance.

%% file: template.bbl
\begin{thebibliography}{44}
\providecommand{\natexlab}[1]{#1}
\providecommand{\url}[1]{\texttt{#1}}
\expandafter\ifx\csname urlstyle\endcsname\relax
  \providecommand{\doi}[1]{doi: #1}\else
  \providecommand{\doi}{doi: \begingroup \urlstyle{rm}\Url}\fi

\bibitem[Hamilton(2023)]{Parsons2023}
Olly Parsons;~Zoe Hamilton.
\newblock Cell broadcast for early warning systems: A review of the technology
  and how to implement it, November 2023.

\bibitem[Al-dalahmeh et~al.(2018)Al-dalahmeh, Al-Shamaileh, Aloudat, and
  Obeidat]{Aldalahmeh2018}
Mahmoud Al-dalahmeh, Ons Al-Shamaileh, Anas Aloudat, and Bader~Yousef Obeidat.
\newblock The viability of mobile services (sms and cell broadcast) in
  emergency management solutions: An exploratory study.
\newblock \emph{International Journal of Interactive Mobile Technologies
  (iJIM)}, 12\penalty0 (1):\penalty0 95, January 2018.
\newblock ISSN 1865-7923.
\newblock \doi{10.3991/ijim.v12i1.7677}.

\bibitem[{NTT DOCOMO Technical Journal Editorial Office}(2012)]{NDTJEO2012}
{NTT DOCOMO Technical Journal Editorial Office}.
\newblock Measures for recovery from the great east japan earthquake using ntt
  docomo r\&d technology.
\newblock \emph{NTT DOCOMO Technical Journal}, 2012.

\bibitem[{ITU}(2010)]{ITU2010}
{ITU}.
\newblock Press release - itu leads effort for reconstruction of haiti’s
  telecommunications and ict infrastructure - secretary general calls for
  broadband access for all inhabitants, July 2010.
\newblock URL
  \url{https://www.itu.int/net/pressoffice/press\_releases/2010/29.aspx}.

\bibitem[Reuters(2017)]{Reuters2017}
Reuters.
\newblock U.{S}. {FCC} says over 95 pct of {Puerto} {Rico} cell sites out of
  service, September 2017.
\newblock URL
  \url{https://www.reuters.com/article/business/environment/us-fcc-says-over-95-pct-of-puerto-rico-cell-sites-out-of-service-idUSKCN1BW2R1/}.

\bibitem[Reuter(2014)]{Reuter2014}
Christian Reuter.
\newblock Communication between power blackout and mobile network overload.
\newblock \emph{International Journal of Information Systems for Crisis
  Response and Management}, 6\penalty0 (2):\penalty0 38--53, April 2014.
\newblock ISSN 1937-9420.
\newblock \doi{10.4018/ijiscram.2014040103}.

\bibitem[Kirimtat et~al.(2020)Kirimtat, Krejcar, Kertesz, and
  Tasgetiren]{Kirimtat2020}
Ayca Kirimtat, Ondrej Krejcar, Attila Kertesz, and M.~Fatih Tasgetiren.
\newblock Future trends and current state of smart city concepts: A survey.
\newblock \emph{IEEE Access}, 8:\penalty0 86448--86467, 2020.
\newblock ISSN 2169-3536.
\newblock \doi{10.1109/access.2020.2992441}.

\bibitem[Zanella et~al.(2014)Zanella, Bui, Castellani, Vangelista, and
  Zorzi]{Zanella2014}
Andrea Zanella, Nicola Bui, Angelo Castellani, Lorenzo Vangelista, and Michele
  Zorzi.
\newblock Internet of things for smart cities.
\newblock \emph{IEEE Internet of Things Journal}, 1\penalty0 (1):\penalty0
  22--32, February 2014.
\newblock ISSN 2372-2541.
\newblock \doi{10.1109/jiot.2014.2306328}.

\bibitem[Ismagilova et~al.(2019)Ismagilova, Hughes, Dwivedi, and
  Raman]{Ismagilova2019}
Elvira Ismagilova, Laurie Hughes, Yogesh~K. Dwivedi, and K.~Ravi Raman.
\newblock Smart cities: Advances in research—an information systems
  perspective.
\newblock \emph{International Journal of Information Management}, 47:\penalty0
  88--100, August 2019.
\newblock ISSN 0268-4012.
\newblock \doi{10.1016/j.ijinfomgt.2019.01.004}.

\bibitem[{European Comission}()]{ComissaoEuropeia_SmartCity}
{European Comission}.
\newblock Smart cities.
\newblock URL
  \url{https://commission.europa.eu/eu-regional-and-urban-development/topics/cities-and-urban-development/city-initiatives/smart-cities\_pt}.

\bibitem[{IoT For All}(2023)]{Beecher_IotForAll_023}
{IoT For All}.
\newblock Building {Resilient} {Smart} {Cities} {\textbar} {Wi}-{SUN}
  {Alliance}'s {Phil} {Beecher}, December 2023.
\newblock URL \url{https://www.youtube.com/watch?v=lGXsdRnVrj0}.

\bibitem[Tian et~al.(2021)Tian, Santi, Seferagić, Lan, and Famaey]{Tian2021}
Le~Tian, Serena Santi, Amina Seferagić, Julong Lan, and Jeroen Famaey.
\newblock Wi-fi halow for the internet of things: An up-to-date survey on ieee
  802.11ah research.
\newblock \emph{Journal of Network and Computer Applications}, 182:\penalty0
  103036, May 2021.
\newblock ISSN 1084-8045.
\newblock \doi{10.1016/j.jnca.2021.103036}.

\bibitem[WFA(2020{\natexlab{a}})]{WFA2020}
Wi-fi halow: Wi-fi for iot applications.
\newblock Technical report, Wi-Fi Alliance, May 2020{\natexlab{a}}.

\bibitem[WFA(2020{\natexlab{b}})]{WFA2021}
Wi-{Fi} {CERTIFIED} {HaLow}™ {Technology} {Overview}.
\newblock Technical report, Wi-Fi Alliance, May 2020{\natexlab{b}}.
\newblock URL
  \url{https://www.wi-fi.org/file/wi-fi-certified-halow-technology-overview-2021}.

\bibitem[Lee et~al.(2021)Lee, Kim, Choi, Park, Lee, Cho, and Yu]{Lee2021}
Il-Gu Lee, Duk~Bai Kim, Jeongki Choi, Hyungu Park, Sok-Kyu Lee, Juphil Cho, and
  Heejung Yu.
\newblock Wifi halow for long-range and low-power internet of things: System on
  chip development and performance evaluation.
\newblock \emph{IEEE Communications Magazine}, 59\penalty0 (7):\penalty0
  101--107, July 2021.
\newblock ISSN 1558-1896.
\newblock \doi{10.1109/mcom.001.2000815}.

\bibitem[Muteba et~al.(2019)Muteba, Djouani, and Olwal]{Muteba2019}
Franck Muteba, Karim Djouani, and Thomas Olwal.
\newblock A comparative survey study on lpwa iot technologies: Design,
  considerations, challenges and solutions.
\newblock \emph{Procedia Computer Science}, 155:\penalty0 636--641, 2019.
\newblock ISSN 1877-0509.
\newblock \doi{10.1016/j.procs.2019.08.090}.

\bibitem[{Open Networking Foundation}()]{ONF_SDN_definition}
{Open Networking Foundation}.
\newblock Software-{Defined} {Networking} ({SDN}) {Definition}.
\newblock URL \url{https://opennetworking.org/sdn-definition/}.

\bibitem[Chilamkurthy et~al.(2022)Chilamkurthy, Pandey, Ghosh, Cenkeramaddi,
  and Dai]{Chilamkurthy2022}
Naga~Srinivasarao Chilamkurthy, Om~Jee Pandey, Anirban Ghosh, Linga~Reddy
  Cenkeramaddi, and Hong-Ning Dai.
\newblock Low-power wide-area networks: A broad overview of its different
  aspects.
\newblock \emph{IEEE Access}, 10:\penalty0 81926--81959, 2022.
\newblock ISSN 2169-3536.
\newblock \doi{10.1109/access.2022.3196182}.

\bibitem[Devalal and Karthikeyan(2018)]{Devalal2018}
Shilpa Devalal and A.~Karthikeyan.
\newblock Lora technology - an overview.
\newblock In \emph{2018 Second International Conference on Electronics,
  Communication and Aerospace Technology (ICECA)}. IEEE, March 2018.
\newblock \doi{10.1109/iceca.2018.8474715}.

\bibitem[Ayoub~Kamal et~al.(2023)Ayoub~Kamal, Alam, Sajak, and
  Mohd~Su’ud]{AyoubKamal2023}
Muhammad Ayoub~Kamal, Muhammad~Mansoor Alam, Aznida Abu~Bakar Sajak, and
  Mazliham Mohd~Su’ud.
\newblock Requirements, deployments, and challenges of lora technology: A
  survey.
\newblock \emph{Computational Intelligence and Neuroscience}, 2023\penalty0
  (1), January 2023.
\newblock ISSN 1687-5273.
\newblock \doi{10.1155/2023/5183062}.

\bibitem[Sallum et~al.(2020)Sallum, Pereira, Alves, and Santos]{Sallum2020}
Eduardo Sallum, Nuno Pereira, Mário Alves, and Max Santos.
\newblock Improving quality-of-service in lora low-power wide-area networks
  through optimized radio resource management.
\newblock \emph{Journal of Sensor and Actuator Networks}, 9\penalty0
  (1):\penalty0 10, February 2020.
\newblock ISSN 2224-2708.
\newblock \doi{10.3390/jsan9010010}.

\bibitem[GAITAN and HOJBOTA(2020)]{GAITAN2020}
Nicoleta~Cristina GAITAN and Paula HOJBOTA.
\newblock Forest fire detection system using lora technology.
\newblock \emph{International Journal of Advanced Computer Science and
  Applications}, 11\penalty0 (5), 2020.
\newblock ISSN 2158-107X.
\newblock \doi{10.14569/ijacsa.2020.0110503}.

\bibitem[Ortigoso et~al.(2024{\natexlab{a}})Ortigoso, Vieira, Fuentes, Frazão,
  Costa, and Pereira]{Ortigoso2024}
Ana~Rita Ortigoso, Gabriel Vieira, Daniel Fuentes, Luis Frazão, Nuno Costa,
  and António Pereira.
\newblock Ddfh: Dynamic dual frequency hopping for lora networks.
\newblock In \emph{2024 34th International Telecommunication Networks and
  Applications Conference (ITNAC)}, pages 1--4. IEEE, November
  2024{\natexlab{a}}.
\newblock \doi{10.1109/itnac62915.2024.10815328}.

\bibitem[Macaraeg et~al.(2020)Macaraeg, Hilario, and Ambatali]{Macaraeg2020}
Khazmir Camille Valerie~G. Macaraeg, Calvin Artemies~G. Hilario, and Charleston
  Dale~C. Ambatali.
\newblock Lora-based mesh network for off-grid emergency communications.
\newblock In \emph{2020 IEEE Global Humanitarian Technology Conference (GHTC)}.
  IEEE, October 2020.
\newblock \doi{10.1109/ghtc46280.2020.9342944}.

\bibitem[Dalpathadu et~al.(2021)Dalpathadu, Showry, Kuppusamy, Udugama, and
  Förster]{Dalpathadu2021}
Yamani Dalpathadu, Thumma Showry, Vishnupriya Kuppusamy, Asanga Udugama, and
  Anna Förster.
\newblock Disseminating data using lora and epidemic forwarding in disaster
  rescue operations.
\newblock In \emph{Proceedings of the Conference on Information Technology for
  Social Good}, GoodIT ’21, pages 293--296. ACM, September 2021.
\newblock \doi{10.1145/3462203.3475917}.

\bibitem[Ranasinghe et~al.(2024)Ranasinghe, Udara, Mathotaarachchi, Thenuwara,
  Dias, Prasanna, Edirisinghe, Gayan, Holden, Punchihewa, Stephens, and
  Drummond]{Ranasinghe2024}
Vinuja Ranasinghe, Nuwan Udara, Movindi Mathotaarachchi, Tharindu Thenuwara,
  Dileeka Dias, Raj Prasanna, Sampath Edirisinghe, Samiru Gayan, Caroline
  Holden, Amal Punchihewa, Max Stephens, and Paul Drummond.
\newblock Rapid and resilient lora leap: A novel multi-hop architecture for
  decentralised earthquake early warning systems.
\newblock \emph{Sensors}, 24\penalty0 (18):\penalty0 5960, September 2024.
\newblock ISSN 1424-8220.
\newblock \doi{10.3390/s24185960}.

\bibitem[Vithayathil et~al.(2021)Vithayathil, Mohamed, Jacob, Raju, and
  Jacob]{Vithayathil2021}
Aishwarya Vithayathil, Afaf Mohamed, Ann Jacob, Chinju~Merin Raju, and Jaison
  Jacob.
\newblock Lora based wireless network for disaster rescue operations.
\newblock In \emph{2021 International Conference on Advances in Computing and
  Communications (ICACC)}, pages 1--7. IEEE, October 2021.
\newblock \doi{10.1109/icacc-202152719.2021.9708218}.

\bibitem[Stiballe et~al.(2023)Stiballe, Welzel, Dorfschmidt, Schluter, Steuter,
  Hense, and Klingler]{Stiballe2023}
Alisa Stiballe, Simon Welzel, Johannes Dorfschmidt, Darvin Schluter,
  Dominik~Delgado Steuter, Jannik~Lukas Hense, and Florian Klingler.
\newblock Demo: Chat based emergency service via long range wireless
  communication (lora).
\newblock In \emph{2023 IEEE 48th Conference on Local Computer Networks (LCN)},
  pages 1--3. IEEE, October 2023.
\newblock \doi{10.1109/lcn58197.2023.10223398}.

\bibitem[Pueyo~Centelles et~al.(2021)Pueyo~Centelles, Meseguer, Freitag,
  Navarro, Ochoa, and Santos]{PueyoCentelles2021}
Roger Pueyo~Centelles, Roc Meseguer, Felix Freitag, Leandro Navarro, Sergio~F.
  Ochoa, and Rodrigo~M. Santos.
\newblock Loramoto: A communication system to provide safety awareness among
  civilians after an earthquake.
\newblock \emph{Future Generation Computer Systems}, 115:\penalty0 150--170,
  February 2021.
\newblock ISSN 0167-739X.
\newblock \doi{10.1016/j.future.2020.07.040}.

\bibitem[Khan et~al.(2017)Khan, Wang, Bhuiyan, and Li]{Khan2017}
Muhammad~Faizan Khan, Guojun Wang, Md~Zakirul~Alam Bhuiyan, and Xu~Li.
\newblock Wi-fi signal coverage distance estimation in collapsed structures.
\newblock In \emph{2017 IEEE International Symposium on Parallel and
  Distributed Processing with Applications and 2017 IEEE International
  Conference on Ubiquitous Computing and Communications (ISPA/IUCC)}. IEEE,
  December 2017.
\newblock \doi{10.1109/ispa/iucc.2017.00162}.

\bibitem[Khan et~al.(2018{\natexlab{a}})Khan, Wang, Bhuiyan, and
  Chen]{Khan2018a}
Muhammad~Faizan Khan, Guojun Wang, Md~Zakirul~Alam Bhuiyan, and Shuhong Chen.
\newblock Wi-fi radar placement for coverage in collapsed structures.
\newblock In \emph{2018 IEEE Intl Conf on Parallel; Distributed Processing with
  Applications, Ubiquitous Computing; Communications, Big Data; Cloud
  Computing, Social Computing; Networking, Sustainable Computing;
  Communications (ISPA/IUCC/BDCloud/SocialCom/SustainCom)}. IEEE, December
  2018{\natexlab{a}}.
\newblock \doi{10.1109/bdcloud.2018.00071}.

\bibitem[Khan et~al.(2018{\natexlab{b}})Khan, Wang, Bhuiyan, and
  Peng]{Khan2018b}
Muhammad~Faizan Khan, Guojun Wang, Md~Zakirul~Alam Bhuiyan, and Tao Peng.
\newblock Wi-fi halow signal coverage estimation in collapsed structures.
\newblock In \emph{2018 IEEE 16th Intl Conf on Dependable, Autonomic and Secure
  Computing, 16th Intl Conf on Pervasive Intelligence and Computing, 4th Intl
  Conf on Big Data Intelligence and Computing and Cyber Science and Technology
  Congress(DASC/PiCom/DataCom/CyberSciTech)}. IEEE, August 2018{\natexlab{b}}.
\newblock \doi{10.1109/dasc/picom/datacom/cyberscitec.2018.00113}.

\bibitem[Khan et~al.(2018{\natexlab{c}})Khan, Wang, Bhuiyan, and
  Xing]{Khan2018}
Muhammad~Faizan Khan, Guojun Wang, Md~Zakirul~Alam Bhuiyan, and Xiaofei Xing.
\newblock Towards wi-fi radar in collapsed structures.
\newblock In \emph{2018 IEEE SmartWorld, Ubiquitous Intelligence; Computing,
  Advanced; Trusted Computing, Scalable Computing; Communications, Cloud; Big
  Data Computing, Internet of People and Smart City Innovation
  (SmartWorld/SCALCOM/UIC/ATC/CBDCom/IOP/SCI)}. IEEE, October
  2018{\natexlab{c}}.
\newblock \doi{10.1109/smartworld.2018.00132}.

\bibitem[Khan et~al.(2019{\natexlab{a}})Khan, Wang, and Bhuiyan]{Khan2019}
Muhammad~Faizan Khan, Guojun Wang, and Md~Zakirul~Alam Bhuiyan.
\newblock Wi-fi frequency selection concept for effective coverage in collapsed
  structures.
\newblock \emph{Future Generation Computer Systems}, 97:\penalty0 409--424,
  August 2019{\natexlab{a}}.
\newblock ISSN 0167-739X.
\newblock \doi{10.1016/j.future.2019.02.061}.

\bibitem[Khan et~al.(2020)Khan, Wang, Bhuiyan, and Yang]{Khan2020}
Muhammad~Faizan Khan, Guojun Wang, Md~Zakirul~Alam Bhuiyan, and Kun Yang.
\newblock Toward wi-fi halow signal coverage modeling in collapsed structures.
\newblock \emph{IEEE Internet of Things Journal}, 7\penalty0 (3):\penalty0
  2181--2196, March 2020.
\newblock ISSN 2372-2541.
\newblock \doi{10.1109/jiot.2019.2959123}.

\bibitem[Khan et~al.(2019{\natexlab{b}})Khan, Wang, and Bhuiyan]{Khan2019a}
Muhammad~Faizan Khan, Guojun Wang, and Md. Zakirul~Alam Bhuiyan.
\newblock Towards debris information analysis and abstraction for wi-fi radar
  edge in collapsed structures.
\newblock \emph{IEEE Access}, 7:\penalty0 168075--168090, 2019{\natexlab{b}}.
\newblock ISSN 2169-3536.
\newblock \doi{10.1109/access.2019.2954281}.

\bibitem[Khorov et~al.(2019)Khorov, Lyakhov, Nasedkin, and Yusupov]{Khorov2019}
E.~M. Khorov, A.~I. Lyakhov, I.~A. Nasedkin, and R.~R. Yusupov.
\newblock Emergency alert delivery in a heterogeneous wi-fi halow network.
\newblock \emph{Journal of Communications Technology and Electronics},
  64\penalty0 (12):\penalty0 1517--1522, December 2019.
\newblock ISSN 1555-6557.
\newblock \doi{10.1134/s1064226919120106}.

\bibitem[Purat et~al.(2022)Purat, Lehmann, Karagulle, and Voisard]{Purat2022}
Justus Purat, Nicolas~J. Lehmann, Muhammed-Ugur Karagulle, and Agnes Voisard.
\newblock Halownet - a wifi halow network-based information system for the
  provision of multi-sided applications for medical emergency scenarios.
\newblock In \emph{2022 IEEE 10th International Conference on Healthcare
  Informatics (ICHI)}, pages 519--521. IEEE, June 2022.
\newblock \doi{10.1109/ichi54592.2022.00095}.

\bibitem[Thangadorai et~al.(2025)Thangadorai, Sivalingam, Pandey, Murugesan,
  and Kanagarathinam]{Thangadorai2025}
Kavin~Kumar Thangadorai, Krishna~M. Sivalingam, Anshul Pandey, Kumar Murugesan,
  and Madhan~Raj Kanagarathinam.
\newblock Wilongh: A custom hand-held platform for long-range halow mesh
  networks in human-to-human communication.
\newblock \emph{IEEE Open Journal of the Communications Society}, pages 1--1,
  2025.
\newblock ISSN 2644-125X.
\newblock \doi{10.1109/ojcoms.2025.3547615}.

\bibitem[Chounos et~al.(2023)Chounos, Maroulis, and Korakis]{Chounos2023}
Kostas Chounos, Manos Maroulis, and Thanasis Korakis.
\newblock On the involvement of ieee 802.11ah enabled unmanned aerial vehicles
  (uavs) in emergency networks.
\newblock In \emph{2023 IEEE Conference on Standards for Communications and
  Networking (CSCN)}, pages 413--416. IEEE, November 2023.
\newblock \doi{10.1109/cscn60443.2023.10453215}.

\bibitem[Riza and Gunawan(2020)]{Riza2020}
Tengku~Ahmad Riza and Dadang Gunawan.
\newblock Ieee 802.11ah network challenges supports covid-19 prevention team.
\newblock In \emph{2020 IEEE 10th International Conference on Electronics
  Information and Emergency Communication (ICEIEC)}, pages 73--76. IEEE, July
  2020.
\newblock \doi{10.1109/iceiec49280.2020.9152346}.

\bibitem[Kane et~al.(2023)Kane, Liu, McKague, and Walker]{Kane2023}
Luke Kane, Vicky Liu, Matthew McKague, and Geoffrey Walker.
\newblock An experimental field comparison of wi-fi halow and lora for the
  smart grid.
\newblock \emph{Sensors}, 23\penalty0 (17):\penalty0 7409, August 2023.
\newblock ISSN 1424-8220.
\newblock \doi{10.3390/s23177409}.

\bibitem[Urbanelli et~al.(2024)Urbanelli, Frisiello, Bruno, and
  Rossi]{Urbanelli2024}
Angelica Urbanelli, Antonella Frisiello, Luca Bruno, and Claudio Rossi.
\newblock The ermes chatbot: A conversational communication tool for improved
  emergency management and disaster risk reduction.
\newblock \emph{International Journal of Disaster Risk Reduction},
  112:\penalty0 104792, October 2024.
\newblock ISSN 2212-4209.
\newblock \doi{10.1016/j.ijdrr.2024.104792}.

\bibitem[Ortigoso et~al.(2024{\natexlab{b}})Ortigoso, Vieira, Fuentes,
  Fraz{\~a}o, Costa, and Pereira]{Ortigoso2024a}
Ana~Rita Ortigoso, Gabriel Vieira, Daniel Fuentes, Luis Fraz{\~a}o, Nuno Costa,
  and António Pereira.
\newblock A multi-tenant sdn architecture for network deployment using a wi-fi
  halow-based ieee 802.11s mesh.
\newblock In \emph{2024 IEEE Virtual Conference on Communications (VCC)}, pages
  1--6. IEEE, December 2024{\natexlab{b}}.
\newblock \doi{10.1109/vcc63113.2024.10914359}.

\end{thebibliography}
